\documentclass{vldb}
\usepackage{balance,array,microtype,pifont,gitinfo,enumitem}

\usepackage[labelfont=bf]{caption}
\usepackage{subcaption}
\usepackage{color, times, mathptmx, amsmath, amssymb, amsthm, verbatim, mdwlist, graphicx, footnote, multirow, xspace, tikz, colortbl,dcolumn, mathtools,
chngcntr}

\usepackage[noadjust]{cite}

\DeclareMathAlphabet{\mathcal}{OMS}{cmsy}{m}{n}
\let\footnotesize\small

\usepackage{algorithm}
\usepackage[noend]{algpseudocode}

\algrenewcommand{\alglinenumber}[1]{\small#1:}

\theoremstyle{definition}
\usepackage{scrextend}

\usepackage[hidelinks]{hyperref}
\usepackage{breakurl}

\newcommand{\minihead}[1]{{\vspace{.45em}\noindent\textbf{#1.} }}

\DeclareMathOperator*{\argmin}{arg\,min}

\newif\ifextended
\extendedtrue

\usepackage{etoolbox}
\newcommand{\zerodisplayskips}{%
  \setlength{\abovedisplayskip}{4pt}
  \setlength{\belowdisplayskip}{4pt}
  \setlength{\abovedisplayshortskip}{4pt}
  \setlength{\belowdisplayshortskip}{4pt}}
\appto{\normalsize}{\zerodisplayskips}
\appto{\small}{\zerodisplayskips}
\appto{\footnotesize}{\zerodisplayskips}

\newfont{\emailaddr}{phvr at 9pt}

\usepackage{inconsolata}

\begin{document}
\conferenceinfo{Proceedings of the VLDB Endowment,}{Vol. 10, No. 11}

\title{ASAP: Prioritizing Attention via Time Series Smoothing}

\author{Kexin Rong, Peter Bailis\\ [1.45mm]    \affaddr{Stanford InfoLab}\\[1mm] {\emailaddr{\{krong, pbailis\}@cs.stanford.edu}}}
\maketitle

\begin{abstract}
Time series visualization of streaming telemetry (i.e., charting of key metrics such as server load over time) is increasingly prevalent in modern data platforms and applications. However, many existing systems simply plot the raw data streams as they arrive, often obscuring large-scale trends due to small-scale noise. We propose an alternative: to better prioritize end users' attention, smooth time series visualizations as much as possible to remove noise, while retaining large-scale structure to highlight significant deviations.
We develop a new analytics operator called ASAP that automatically smooths streaming time series by adaptively optimizing the trade-off between noise reduction (i.e., variance) and trend retention (i.e., kurtosis).
We introduce metrics to quantitatively assess the quality of smoothed plots and provide an efficient search strategy for optimizing these metrics that combines techniques from stream processing, user interface design, and signal processing via autocorrelation-based pruning, pixel-aware preaggregation, and on-demand refresh.
We demonstrate that ASAP can improve users' accuracy in identifying long-term deviations in time series by up to 38.4\% while reducing response times by up to 44.3\%. Moreover, ASAP delivers these results several orders of magnitude faster than alternative search strategies.
 \end{abstract}

\section{Introduction}
\label{sec:intro}

Data volumes continue to rise, fueled in large part by an increasing
number of automated sources, including sensors, processes, and
devices. For example, each of LinkedIn, Twitter, and Facebook reports
that their production infrastructure generates over 12M events per
second~\cite{twitter-volume, linkedin-volume, fb-volume}.  As a
result, the past several years have seen an explosion in the
development of platforms for managing, storing, and querying
large-scale data streams of time-stamped data---i.e., time
series---from on-premises databases including
InfluxDB~\cite{InfluxDB}, Ganglia~\cite{Ganglia},
Graphite~\cite{Graphite}, OpenTSDB~\cite{OpenTSDB},
Prometheus~\cite{Prometheus}, and Facebook Gorilla~\cite{fb-volume},
to cloud services including DataDog~\cite{Datadog}, New
Relic~\cite{newrelic}, AWS CloudWatch~\cite{cloudwatch}, Google
Stackdriver~\cite{stackdriver}, and Microsoft Azure
Monitor~\cite{azure}. These time series engines provide application authors, site
operators, and ``DevOps'' engineers a means of performing monitoring,
health checks, alerting, and analysis of unusual events such as
failures~\cite{devops,sre}.

While these engines have automated and optimized common tasks in the
storage and processing of time series, effective visualization of time
series remains a challenge. Specifically, in conversations with
\hyphenation{DevOps} engineers using time series data and databases in
cloud services, social networking, industrial manufacturing,
electrical utilities, and mobile applications, we learned that many
production time series visualizations (i.e., ``dashboards'') simply
display raw data streams as they arrive. Engineers reported this
display of raw data can be a poor match for production scenarios
involving data exploration and debugging. That is, as data arrives in
increasing volumes, even small-scale fluctuations in data values can
obscure overall trends and behavior.
For example, an electrical utility employs two staff to perform 24-hour
monitoring of generators. It is critical
that these staff quickly identify any systematic shifts of generator metrics in their monitoring dashboards, even those that are ``sub-threshold'' with respect to a critical alarm. Unfortunately, such sub-threshold events are easily
obscured by short-term fluctuations in the visualization.

\begin{figure}
\includegraphics[width=\linewidth]{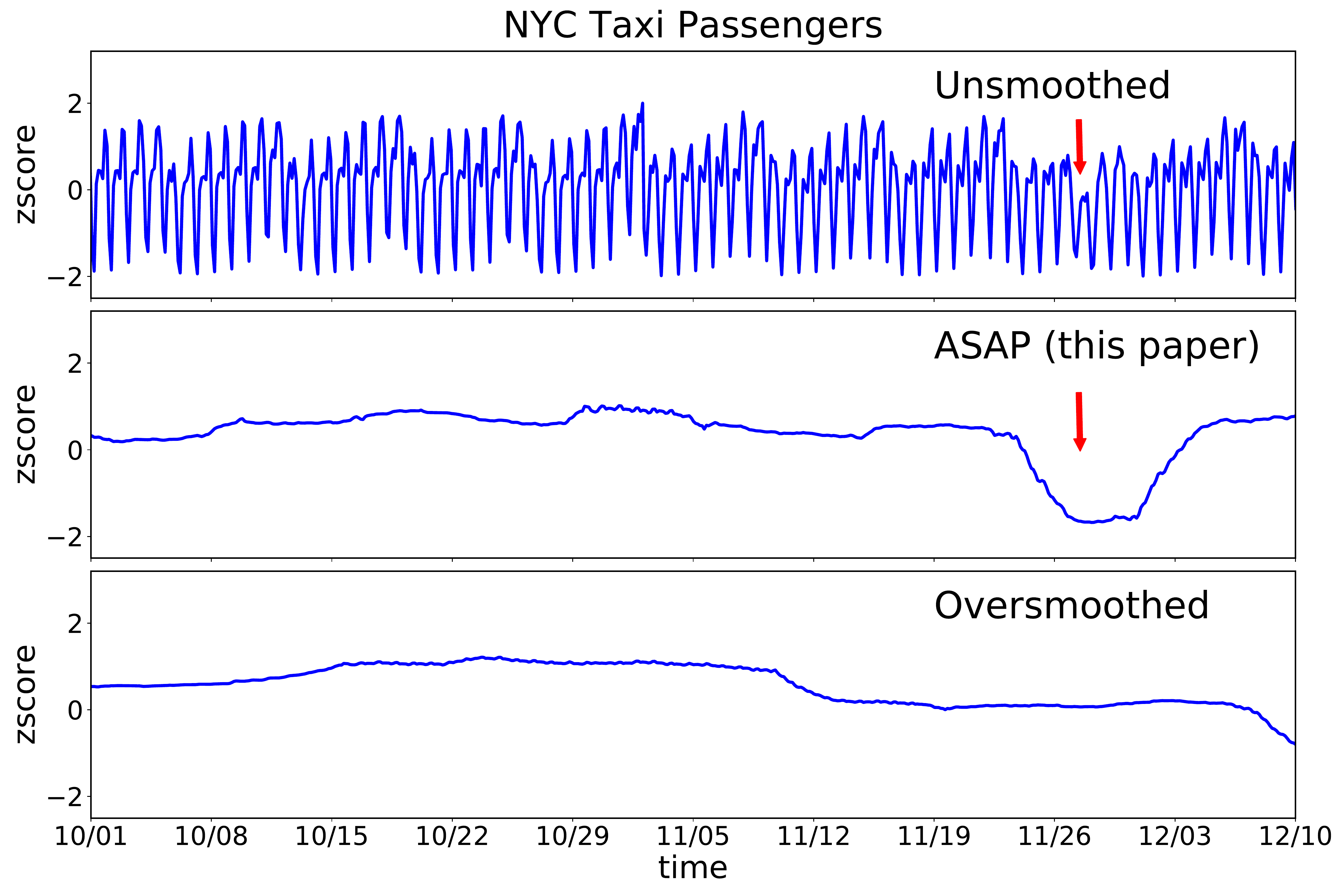}
\caption[]{Normalized number of NYC taxi passengers over 10 weeks.\footnotemark~From top to bottom, the three plots show the hourly average (unsmoothed), the weekly average (smoothed) and the monthly average (oversmoothed) of the same time series. The arrows point to the week of Thanksgiving (11/27), when the number of passengers dips. This phenomenon is most prominent in the smoothed plot produced by ASAP, the subject of this paper.}
\label{fig:taxi}
\end{figure}
\footnotetext{Here and later in this paper, we depict
  z-scores~\cite{zscore} instead of raw values. This choice of
  visualization provides a means of normalizing the visual field
  across plots while still highlighting large-scale trends.}

The resulting challenge in time series visualization at scale is presenting the \textit{appropriate} plot that prioritizes users' attention towards significant deviations. To illustrate this challenge using public data,
consider the time series depicted in Figure~\ref{fig:taxi}. The top plot shows raw data: an hourly average of the
number of NYC taxi passengers over 75 days in 2014~\cite{Numenta}. Daily
fluctuations of taxi volume dominate the visual field, obscuring a significant long-term deviation: the number of taxi passengers experienced a sustained dip
during the week of Thanksgiving. Ideally, we would smooth the local
fluctuations to highlight this deviation in the visualization
(Figure~\ref{fig:taxi}, middle). However, if we smooth too
aggressively, the visualization may hide this trend entirely (Figure~\ref{fig:taxi},
bottom).


In this paper, we address the challenge of prioritizing attention in
time series using a simple strategy: smooth time series visualizations
as much as possible while preserving large-scale deviations. This
raises two key questions. First, how can we quantitatively assess the
quality of a given visualization in removing small-scale variations
and highlighting significant deviations?  Second, how can we use such
a quantitative metric to produce high-quality visualizations quickly
and at scale? We answer both questions through the design of a new
time series visualization operator, called ASAP (Automatic Smoothing
for Attention Prioritization), which quantifiably improves end-user
accuracy and speed in identifying significant deviations in time
series, and is optimized to execute at scale.\footnote{Demo and code
  available at \url{http://futuredata.stanford.edu/asap/}}

To address the first question of quantitative metrics for prioritizing
attention, we combine two statistics. First, we quantify the smoothness
of a time series visualization via the \emph{variance of first
  differences}~\cite{firstdiff}, or the variation of difference between consecutive
points in the series. By applying a moving average of increasing length,
we can reduce this variance and smooth the plot. However, as
illustrated by Figure~\ref{fig:taxi}, it is possible to oversmooth and
obscure the trend entirely. Therefore, to prevent oversmoothing, we
introduce a constraint based on preserving the
\emph{kurtosis}~\cite{kurtosismeaning}---a measure of the
``outlyingness'' of a distribution---of the original time series, 
preserving its structure. Incidentally, this kurtosis measure can
also determine when \emph{not} to smooth (e.g., if a series has a few
well-defined outlying regions). We demonstrate the utility of this
combination of smoothness measure and constraint via two
user studies: compared to displaying raw data, smoothing
time series visualizations using these metrics improves users' accuracy in
identifying anomalies by up to 38.4\% and decreases
response times by up to 44.3\%.

Using these metrics, ASAP automatically chooses smoothing parameters
on users' behalf, producing the smoothest visualization that retains
large-scale deviations.  Given a window of time to visualize (e.g.,
the past 30 minutes of a time series), ASAP selects and applies an
appropriate smoothing parameter to the target series. Unlike existing
smoothing techniques that are designed to produce visually
indistinguishable representations of the original signal (e.g.,
~\cite{M4,linesimpeval}), ASAP is designed to ``distort''
visualizations (e.g. by removing local fluctuations) to highlight key
deviations (e.g. as in Figure~\ref{fig:taxi}) and prioritize end user
attention~\cite{mb-cidr}.

There are three main challenges in enabling this efficient, automatic
smoothing. First, our target workloads exhibit large data volumes---up to millions of events per second---so ASAP must produce legible visualizations despite high volume. Second, to support interactive use, ASAP must render quickly. As we demonstrate, an exhaustive search over smoothing parameters for 1M points requires over an hour, yet we target sub-second response times. Third, appropriate smoothing parameters may change over time: a high-quality parameter choice for one time period may oversmooth or undersmooth in another. Therefore, ASAP must adapt its smoothing parameters in response to changes in the streaming time series.

To address these challenges, ASAP combines techniques from stream processing, user interface design, and signal processing.
First, to scale to large volumes, ASAP pushes constraints regarding the target end-user display into its design. ASAP exploits the fact that its results are designed to be displayed in a fixed number of pixels (e.g., maximum 1334 pixels at a time on the iPhone 7), and uses target resolution as a natural lower bound for the parameter search: there is rarely benefit in choosing parameters that would result in a resolution greater than the target display size. Accordingly, ASAP pre-aggregates data, thus reducing the search space. Second, to further improve rendering time, ASAP prunes the search space by searching for period-aligned time windows (i.e., time lag with high autocorrelation) for periodic data and performing binary search for aperiodic data; we demonstrate both analytically and empirically that this search strategy leads to smooth aggregated series. 
Third, to quickly respond to changes in fast-moving time series, ASAP avoids recomputing smoothing parameters from scratch upon the arrival of each new data point. Instead, ASAP reuses computation and re-renders visualizations on human-observable timescales.

In total, ASAP achieves its goals of efficient and automatic smoothing by treating visualization properties including end-user display constraints and limitations of human perception as critical design considerations. As we empirically demonstrate, this co-design yields useful results, quickly and without manual tuning. We have implemented ASAP as a time series explanation operator in the MacroBase fast data engine~\cite{MB}, and as a Javascript library. The resulting ASAP prototypes demonstrate order-of-magnitude runtime improvements over alternative search strategies while producing high-quality smoothed visualizations.

In summary, we make the following contributions in this work:
\begin{itemize}[itemsep=0.1em, topsep=0.25em]
\item ASAP, the first stream processing operator for
  automatically smoothing time series to reduce local variance (i.e., minimize roughness) while
  preserving large-scale deviations (i.e., preserving kurtosis) in visualization.
\item Three optimizations for improving ASAP's execution speed that leverage
  $i)$ target device resolution in pre-aggregation, $ii)$
  autocorrelation to exploit periodicity, $iii)$ and partial
  materialization for streaming updates.
\item A quantitative evaluation demonstrating ASAP's ability to
  improve user accuracy and response time and deliver
  order-of-magnitude performance improvements.
\end{itemize}

The remainder of this paper proceeds as follows. Section~\ref{sec:background} describes ASAP's architecture and provides additional background regarding our target use cases. Section~\ref{sec:problem} formally introduces ASAP's problem definition and quantitative target metrics. Section~\ref{sec:asap} presents ASAP's search strategy, optimizations, and streaming execution mode. Section~\ref{sec:eval} evaluates ASAP's visualization quality through two user studies, and ASAP's performance on a range of synthetic and real-world time series. Section~\ref{sec:relatedwork} discusses related work, and Section~\ref{sec:conclusion} concludes.

\section{Architecture and Usage}
\label{sec:background}

ASAP provides analysts and system operators an effective and
efficient means of highlighting large-scale deviations
in time series visualizations. In this section, we describe ASAP's
usage and architecture, illustrated via two additional case studies.

Given an input time series (i.e., set of temporally ordered data
points) and target interval for visualization (e.g., the last twelve
hours of data), ASAP returns a transformed, smoothed time series
(e.g., also of twelve hours, but with a smoothing function applied)
for visualization. In the streaming setting, as new data points
arrive, ASAP continuously smooths each fixed-size time interval,
producing a sequence of smoothed time series. Thus, ASAP acts as a
transformation over fixed-size sliding windows over a single time
series. When ASAP users change the range of time series to visualize
(e.g., via zoom-in, zoom-out, scrolling), ASAP re-renders its output
in accordance with the new range. For efficiency, ASAP also allows
users to specify a target display resolution (in pixels) and a desired
refresh rate (in seconds).

ASAP can run either client-side or server-side. For easy integration
with web-based front-ends, ASAP can execute on the client; we provide
a JavaScript library for doing so. However, for resource-constrained
clients, or for servers with a large number of visualization
consumers, ASAP can execute on the server, sending clients the
smoothed stream; this is the execution mode that
MacroBase~\cite{MB} adopts, and MacroBase's ASAP
implementation is portable to existing stream processing engines.

ASAP acts as a modular tool in time series visualization. It can
ingest and process raw data from time series databases such as
InfluxDB, as well as from visualization clients such as plotting
libraries and frontends. For example, when building a monitoring
dashboard, a DevOps engineer could employ ASAP and plot the smoothed
results in his metrics console, or, alternatively, overlay the
smoothed plot on top of the original time series. ASAP can also
\emph{post-process} outputs of time series analyses including motif
discovery, anomaly detection, and clustering~\cite{KeoghEEG, Keoghsin,
  burst, WarrenLiao20051857}: given a single time series as output
from each of these analyses, ASAP can smooth the time series prior to
visualization.

\begin{figure}
\center
\includegraphics[width=0.96\linewidth]{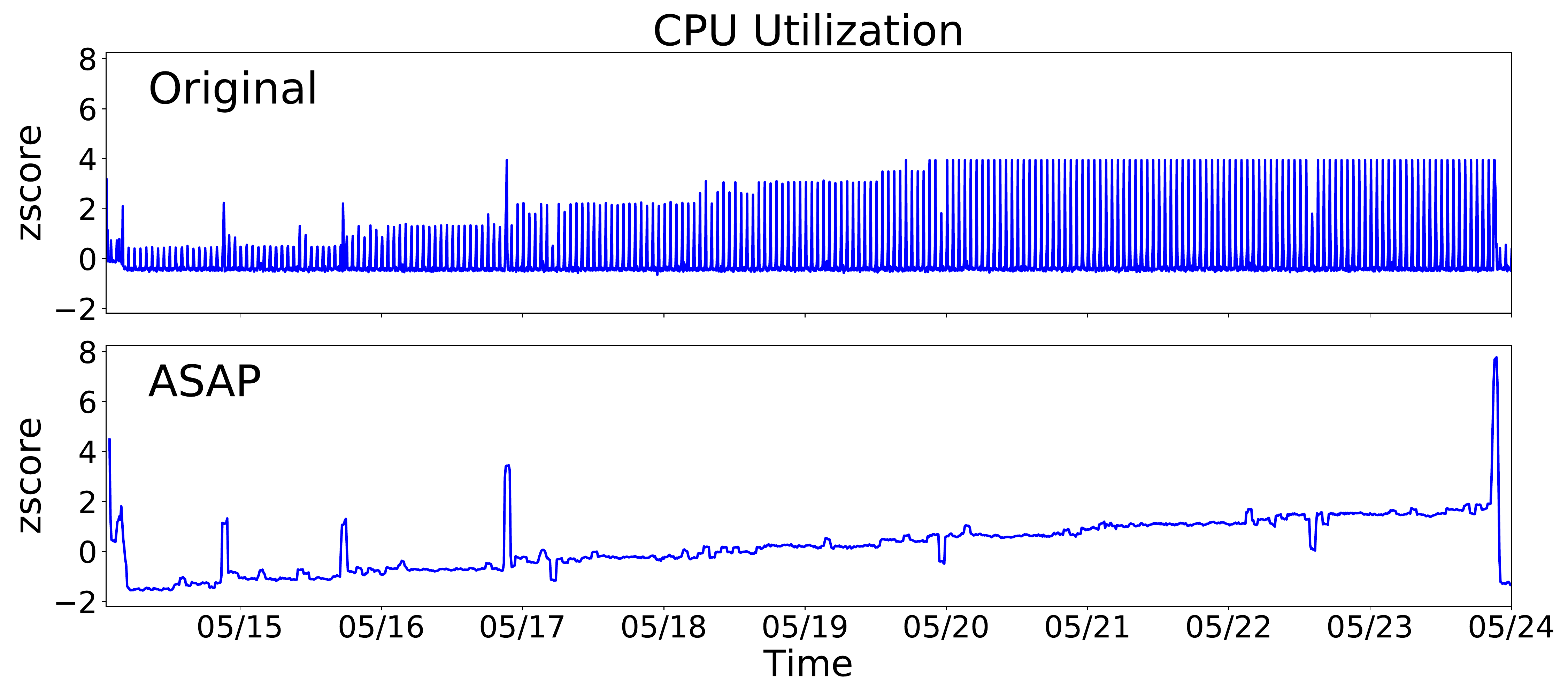}
\vspace{-0.5em}
\caption{Server CPU usage across a cluster over ten days~\cite{Numenta}, visualized via a 5 minute average (raw) and an hourly average (via ASAP). The CPU usage spike around May 24th is obscured by frequent fluctuations in the raw time series.}
\label{fig:cpu}
\end{figure}

To further illustrate ASAP's potential uses in prioritizing attention in time series, we provide two additional case studies cases below, and additional examples of raw time series and their smoothed counterparts in our extended Technical Report (\cite{asaponline}, Appendix~C):

\minihead{Application Monitoring} An on-call application operator is paged at 4AM due to an Amazon CloudWatch alarm on a sudden increase in CPU utilization on her Amazon Web Service cloud instances. After reading the alert message, she accesses her cluster telemetry plots that include CPU usage over the past ten days on her smartphone to obtain a basic understanding of the situation. However, the smartphone's display resolution is too small to effectively display all 4000 readings; as a result, the lines are closely stacked together in the plot, making CPU usage appear stable (Figure \ref{fig:cpu}, top).\footnote{This plot is inspired by an actual use case we encountered in production time series from a large cloud operator; high frequency fluctuations in the plot made it appear that a server was behaving abnormally, when in fact, its overall (smoothed) behavior was similar to others in the cluster.} Unable to obtain useful insights from the plot, the operator must rise from bed and begin checking server logs manually to diagnose the issue. If she were to instead apply ASAP, the usage spike around May 24th would no longer be hidden by noise. 

\begin{figure}
\center
\includegraphics[width=0.95\linewidth]{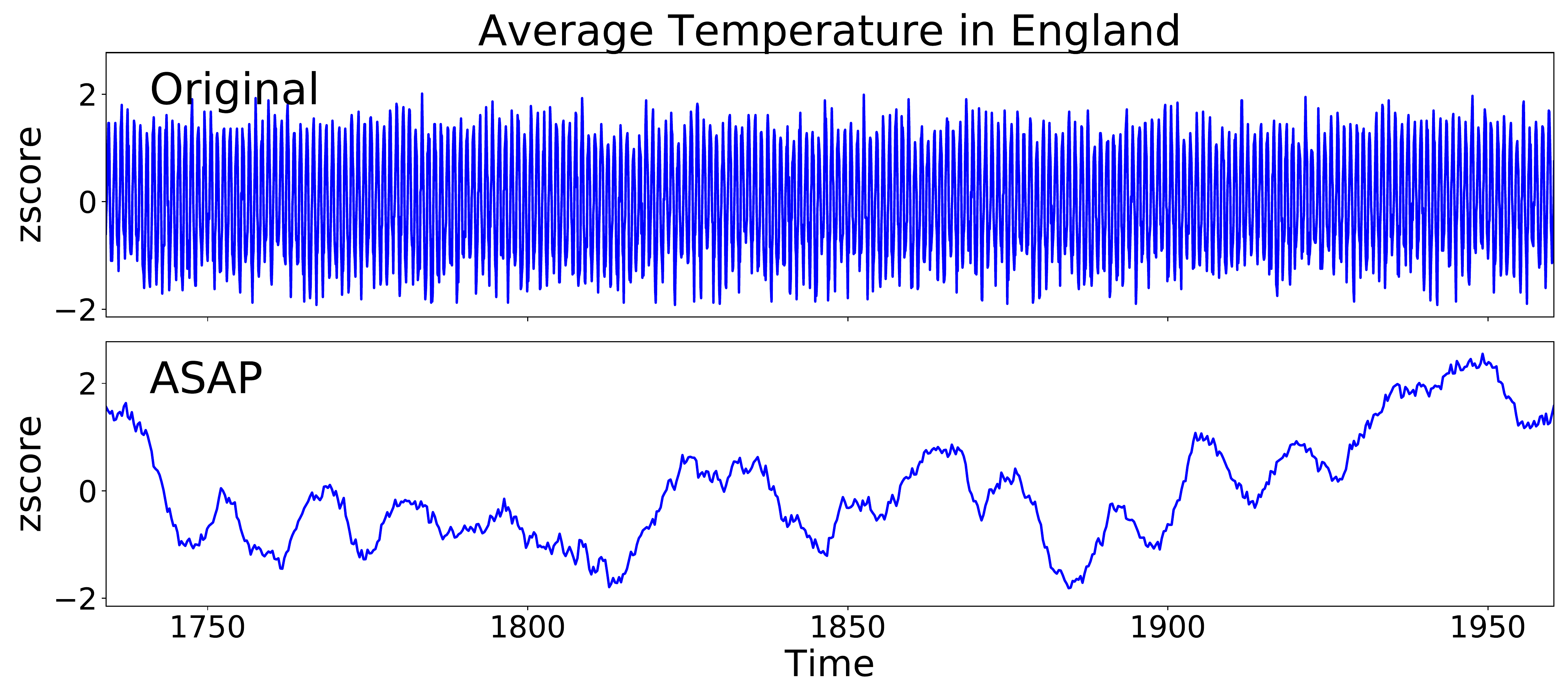}
\vspace{-0.5em}
\caption{Temperature in England from 1723 to 1970~\cite{TSDL}, visualized via a monthly average (raw) and 23-year average (via ASAP). Fluctuations in the raw time series obscure the overall trend.}
\label{fig:temp}
\end{figure}

\minihead{Historical Analyses} A researcher interested in climate change examines a data set of monthly temperature in England over 200 years. When she initially plots the data to determine long-term trends, her plot spills over five lengths of her laptop screen.\footnote{This is not a theoretical example; in fact, the site from which we obtained this data~\cite{TSDL} plots the time series in a six-page PDF. This presentation mode captures fine-grained structure but makes it difficult to determine long-term trends at a glance, as in Figure~\ref{fig:temp}.} Instead of having to scroll to compare temperature in the 1700s with the 1930s, she decides to plot the data herself to fit the entire time series onto one screen. Now, in the re-plotted data (Figure~\ref{fig:temp}, top), seasonal fluctuations each year obscure the overall trend. Instead, if she were to instead use ASAP, she would see a clear trend of rising temperature in the 1900s (Figure~\ref{fig:temp}, bottom).

\section{Problem Definition}
\label{sec:problem}
In this section, we introduce the two key metrics that ASAP uses to
assess the quality of smoothed visualizations as well as its smoothing function. We subsequently cast ASAP's parameter search as an optimization problem.

\subsection{\hspace{-0.2em}Roughness Measure}
As we have discussed, noise and/or frequent fluctuations can distract users from identifying large-scale trends in time series visualizations. Therefore, to prioritize user attention, we wish to \emph{smooth as much as possible while preserving systematic deviations}. We first introduce a metric to quantify the degree of smoothing.

\begin{figure}
\center
\includegraphics[width=0.8\linewidth]{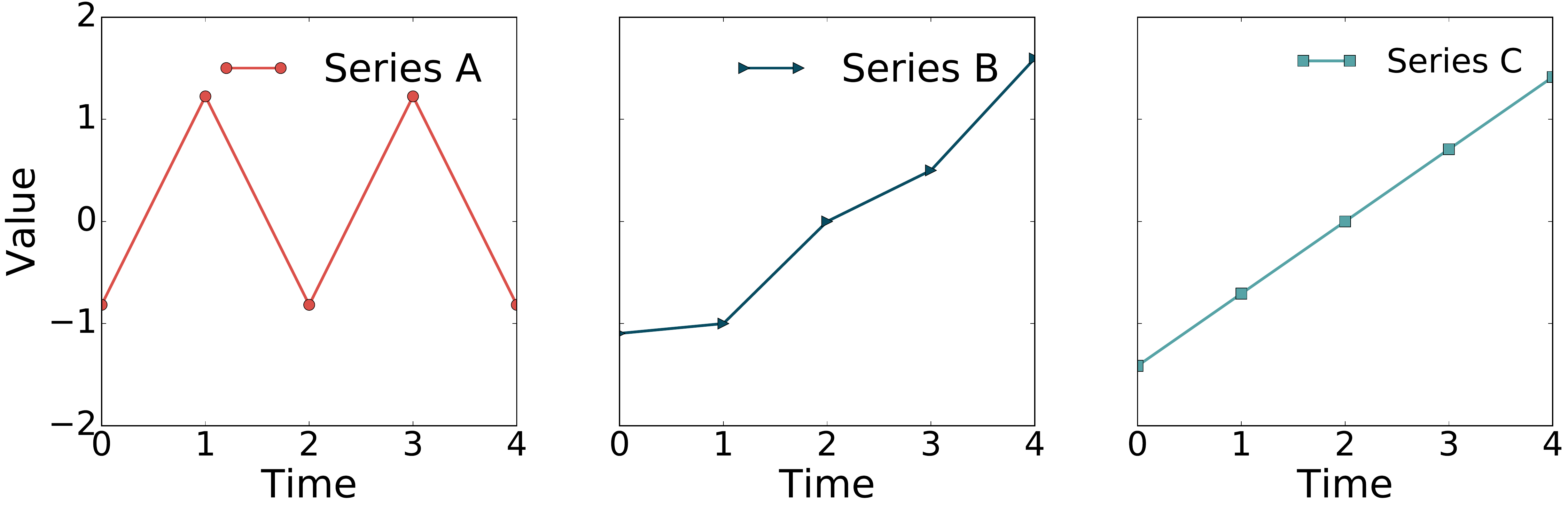}\vspace{-.5em}
\caption{Three time series that appear visually distinct yet all have mean of zero and standard deviation of one. This example illustrates that standard summary statistics such as mean and standard deviation can fail to capture the visual ``smoothness'' of time series.}
\label{fig:roughness}
\end{figure}

Standard summary statistics such as mean and standard deviation alone may not suffice to capture a time series's visual smoothness. For example, consider the three time series in Figure~\ref{fig:roughness}: a jagged line (series A), a slightly bent line (series B), and a straight line (series C). These time series appear different, yet all have a mean of zero and standard deviation of one. However, series C looks ``smoother'' than series A and series B because it has a constant slope. Put another way, the differences between consecutive points in series C have smaller variation than consecutive points in series A and B.

To formalize this intuition, we define the roughness (i.e., inverse ``smoothness,'' to be minimized) of a time series as the standard deviation of the differences between consecutive points in the series. The smaller the variation of the differences, the smoother the time series.
Formally, given time series $X = \{x_1, x_2, ..., x_N\}$, $x_i \in \mathbb{R}$, we adopt the concept of the \textit{first difference} series~\cite{firstdiff} as:
\[\Delta X = \{\Delta x_1, \Delta x_2, ...\}\quad s.t.\quad \Delta x_i = x_{i + 1} - x_i, i \in \{1, 2, ..., N - 1\}\]
Subsequently, we can define the roughness of time series $X$ as the standard deviation of the first difference series:
\[\mathrm{roughness}(X) = \sigma(\Delta X)\]
This use of variance of differences is closely related to the concept of a \emph{variogram}~\cite{variogram}, a commonly-used measure in spatial statistics (especially geostatistics) that characterizes the spatial continuity (or \textit{surface roughness}) of a given dataset. By this definition, the roughness of the three time series in Figure~\ref{fig:roughness} are 2.04, 0.4, and 0, respectively. Note that a time series will have roughness value of 0 if and only if the corresponding plot is a straight line (like series C). Specifically, a roughness value of 0 implies the differences between neighboring points are identical and therefore the plot corresponding to the series will have a constant slope, resulting in a straight line.

\subsection{\hspace{-0.2em}Preservation Measure}
Per the above observation, if we simply minimize roughness, we will produce plots that approximate straight lines. In some cases, this is desirable; if the overall trend is a straight line, then removing noise may, in fact, result in a straight line. However, as our examples in Section~\ref{sec:background} demonstrate, many meaningful trends are not accurately represented by straight lines. As a result, we need a measure of ``trend preservation'' that captures how well we are preserving large-scale deviations within the time series.

To quantify how well we are preserving large deviations in the original time series, we measure the distribution \textit{kurtosis}~\cite{kurtosismeaning}. Kurtosis captures ``tailedness'' of the probability distribution of a real-valued random variable, or how much mass is near the tails of the distribution. More formally, given a random variable $X$ with mean $\mu$ and standard deviation $\sigma$, kurtosis is defined as the fourth standardized moment:   
\[ \mathrm{Kurt}[X] = \frac{\mathbb{E}[(X - \mu)^4]}{\mathbb{E}[(X - \mu)^2]^2}\]

\begin{figure}
\includegraphics[width=\columnwidth]{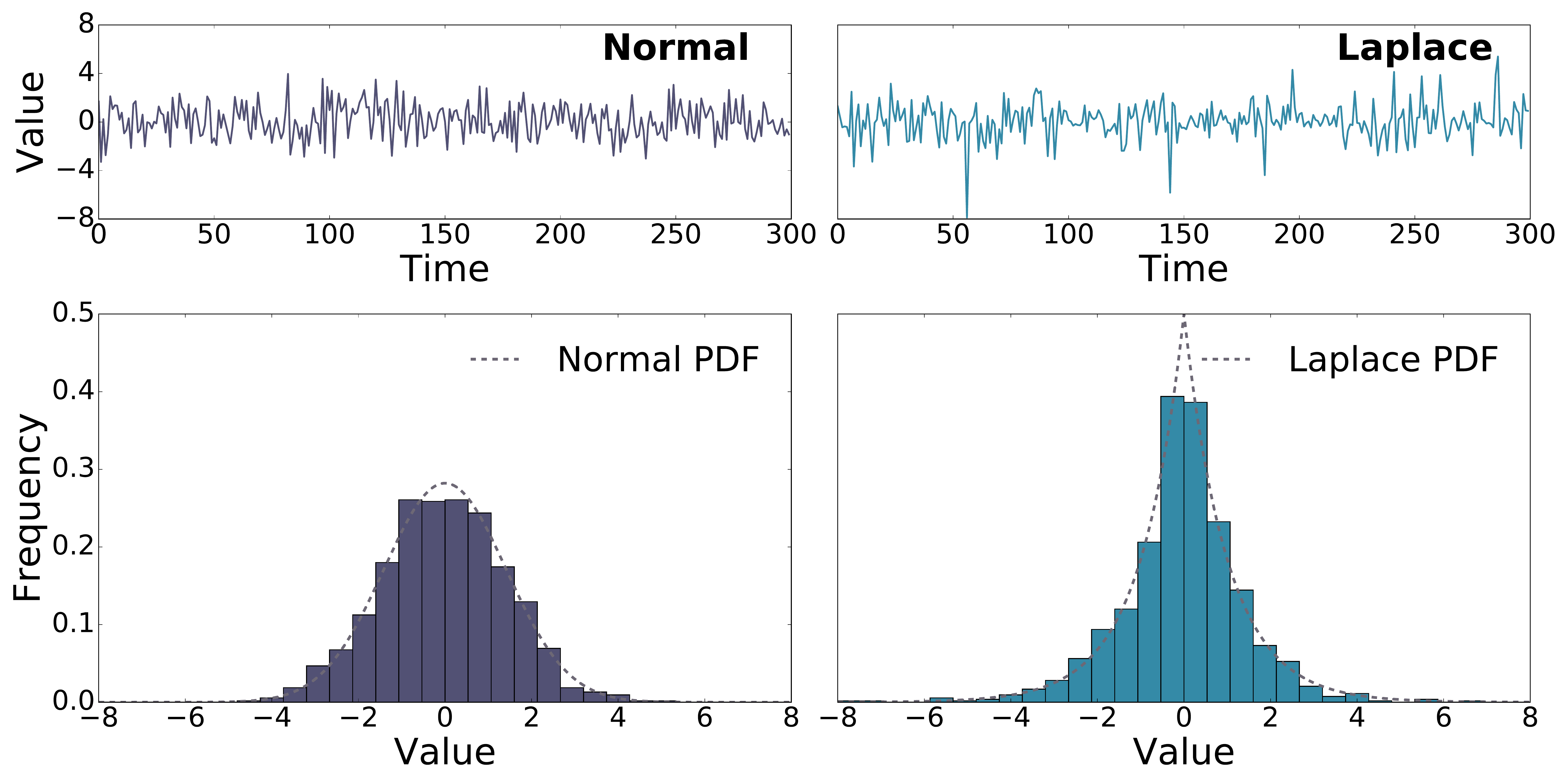}
\caption{Time series and histograms sampled from a normal distribution (left) and a Laplace distribution (right). Despite having the same mean (0) and variance (2), the Laplace series includes a few large deviations, while the normal includes a large number of moderate deviations. The difference in tendency to produce outliers is captured by kurtosis: the normal distribution has a kurtosis of 3, while the Laplace distribution has a kurtosis of 6.}
\label{fig:kurtexample}
\end{figure}

Higher kurtosis means that more of the variance is contributed by rare and extreme deviations, instead of more frequent and modestly sized deviations~\cite{westfall2014kurtosis}. For reference, the kurtosis for univariate normal distribution is 3. Distributions with kurtosis less than 3, such as the uniform distribution, produce fewer and less extreme outliers compared to normal distributions. Distributions with kurtosis larger than 3, such as the Laplace distribution, have heavier tails compared to normal distributions. Figure~\ref{fig:kurtexample} illustrates two time series sampled from the normal and Laplace distribution discussed above. Despite having the same mean and variance, kurtosis captures the two series' difference in tendency to produce outliers.

To prevent oversmoothing large-scale deviations in the original time series, we compare the kurtosis of the time series before and after applying the smoothing function. If the kurtosis of the original series is greater than or equal to the smoothed series, then the proportion of values that significantly deviate in the smoothed series is no smaller than the proportion in the original series. If smoothing is effective, then the smoothing will ``concentrate'' the values around regions of large deviation (i.e., significant shifts from the mean) and therefore highlight these deviations.

If the original time series only contains a few extreme outliers, the smoothing is likely to only average out the deviations, which we also account for in our parameter selection procedure. For example, consider a time series with all but one point in the range $[-1, 1]$ and a single outlying point that has a value of 10. 
This outlier may be the most important piece of information that users would like to highlight in the time series, so applying a simple moving average only decreases the extent of this deviation (i.e., the kurtosis of the smoothed time series decreases). The kurtosis preservation constraint thus ensures we leave the original time series unsmoothed. 

\subsection{\hspace{-0.2em}Smoothing Function}
Given our roughness and preservation measure, we wish to smooth our time series as much as possible (i.e., minimizing roughness) while preserving large-scale deviations (i.e., preserving kurtosis). To perform the actual smoothing, we need a smoothing function.

In this paper, we focus on simple moving average (SMA) as the smoothing function. Three reasons motivate this choice. First, SMA is well studied in the stream processing literature, with several existing techniques for efficient execution and incremental maintenance~\cite{panes}. We adopt these techniques, while using roughness and preservation metrics as a means of automatically tuning SMA parameters for visual effect. Second, SMA is also well studied in the signal processing community. Statistically, the moving average is optimal for recovering the underlying trend of the time series when the fluctuations about the trend are normally distributed~\cite{Smith2003277}, despite its light computational footprint and conceptual simplicity compared to alternatives. Third, we experimented with several alternatives including the MinMax aggregation, the Fourier transform~\cite{STFT}, and the Savitzky-Golay filter~\cite{SG}; SMA had fewer parameters to tune and proved more effective at smoothing per our target metrics. We include a visual comparison in~\cite{asaponline}.

Given input $w \in \mathbb{N}$, SMA averages every sequential set of $w$ points in the original time series $X$ to produce one point in the smoothed series $Y$. 

We can express SMA as:
\vspace{-0.5em}
\[\mathrm{SMA}(X, w) = \{y_1, ... y_{N-w}\} \quad s.t. \quad y_i = \frac{1}{w} \sum_{j = 0}^{w - 1} x_{i + j}\]
\vspace{-0.5em}

When applying SMA over data streams with a sliding window, users can adjust its \textit{window size} (number of points in each window) and \textit{slide size} (distance between neighboring windows) parameter.
In time series visualization, slide size determines the sampling frequency of the original time series and, therefore, the number of distinct, discrete data points in the smoothed plot. In this work, we focus on automatically selecting a window size for a given slide size. Instead of tuning slide size, we employ a policy that sets slide size according to the desired number of points (i.e., pixels) in the final visualization (i.e., $\frac{\textrm{\# original points}}{\textrm{\#desired points}}$). Increasing the slide size beyond this threshold results in fewer data points than specified in the smoothed visualization, and decreasing the slide sizes results in a smoothed time series with more data points than available display resolution. Therefore, we found that varying the slide size did not dramatically improve visualization quality.

\subsection{\hspace{-0.2em}ASAP Problem Statement}
Given our roughness and preservation measures and smoothing function, we present ASAP's problem statement as follows:\\[0mm] 

\noindent\textit{\textsc{Problem.} Given time series $X = \{x_1, x_2, ..., x_N\}$, let $Y = \{y_1, y_2, ...,$\\ $y_{N-W}\}$ be the smoothed series of $X$ obtained by applying a simple moving average with window size $w$ (i.e., $y_i = \frac{1}{w} \sum_{j = 0}^{w - 1} x_{i + j}$). 
Find window size $\hat{w}$ where:
\[ \hat{w} = \argmin_{w} \sigma(\Delta(Y)) \quad s.t.\quad  \mathrm{Kurt}[Y] \geq \mathrm{Kurt}[X]\]}
That is, we wish to reduce roughness in a given time series as much as possible by applying a sliding window average function to the data while preserving kurtosis. 

\section{ASAP}
\label{sec:asap}
In this section, we describe ASAP's core search strategy and optimizations for solving the problem of smoothing parameter selection. We first focus on smoothing a single, fixed-length time series, beginning with a walkthrough of a strawman solution (Section~\ref{asap:strawman}). We then analyze the problem dynamics under a simple, IID distribution (Section~\ref{sec:iid}) and, using the insights from this analysis, we develop a pruning optimization based on autocorrelation (Section~\ref{asap:ac}). We then introduce a pixel-aware optimization that greatly reduces the input space via preaggregation (Section~\ref{sec:pixel}). Finally, we discuss the streaming setting (Section~\ref{sec:stream}).

\subsection{\hspace{-0.2em}Strawman Solution}
\label{asap:strawman}
As a strawman solution, we could exhaustively search all possible window lengths and return the one that gives the smallest roughness measure while satisfying the kurtosis constraint. For each candidate window length, we need to smooth the series and evaluate the roughness and kurtosis. Each of these computations requires linear time ($O(N)$). However, there are also many candidates to evaluate: for a time series of size $N$, we may need to evaluate up to $N$ possible window lengths, resulting in a total running time of $O(N^2)$. As we illustrate empirically in Section~\ref{sec:eval}, in the regime where $N$ is even modestly large, this computation can be prohibitively expensive.

We might consider improving the runtime of this exhaustive search by performing grid search via a sequence of larger step sizes, or by performing binary search. However, as we will demonstrate momentarily, the roughness metric is not guaranteed to be monotonic in window length and therefore, the above search strategies may deliver poor quality results. 
Instead, in the remainder of this section, we describe an alternative search strategy that is able to retain the quality of exhaustive search while achieving meaningful speedups by quickly pruning unpromising candidates and by optimizing for the desired pixel density.

\subsection{\hspace{-0.2em}Basic IID Analysis}
\label{sec:iid}
To leverage properties of the roughness metric to speed ASAP's search, we first consider how window length affects the roughness and kurtosis of the smoothed series.

As a first step, consider a time series $X: \{x_1, x_2, ..., x_N\}$ consisting of samples drawn identically independently distributed (IID) from some distribution with mean $\mu$ and standard deviation $\sigma$. After applying a moving average of window length $w$, we obtain the smoothed series:
\[Y = \mathrm{SMA}(X, w), \quad y_i = \frac{1}{w} \sum_{j = 0}^{w - 1} x_{i + j}, i \in \{1, 2, ..., N-w\}\]

\noindent We denote the first difference series as $\Delta Y = \{\Delta y_1, \Delta y_2, ...\}$, where
\[ \Delta y_i = y_{i + 1} - y_{i} = \frac{1}{w} \sum_{j = 0}^{w - 1}(x_{i + j + 1} - x_{i + j}) = \frac{1}{w}(x_{i + w} - x_i)\]
For convenience, we also denote the first $N-w$ points of $X$ as $X_f = \{x_1, x_2, ..., x_{N-w}\}$ and the last $N-w$ points of $X$ as $X_l = \{x_{w + 1},\\$ $x_{w+2}, ..., x_{N}\}$. Then $\Delta Y = \frac{1}{w}(X_l - X_f)$, and roughness of the smoothed series $Y$ can be written as: 
\begin{equation}
\mathrm{roughness}(Y) = \sigma(\Delta Y) = \frac{1}{w}\sqrt{\mathrm{var}(X_f) + \mathrm{var}(X_l) - 2\mathrm{cov}(X_l, X_f)}
\label{eq:roughness}
\end{equation}
Since each $x_i$ is drawn IID from the same distribution, we have $\mathrm{var}(X_f) = \mathrm{var}(X_l) = \sigma^2$ and $\quad \mathrm{cov}(X_f, X_l) = 0$. Substituting in Equation~\ref{eq:roughness} we obtain: 
\begin{equation}
\label{eq:roughnessiid}
\mathrm{roughness}(Y) = \frac{\sqrt{2}\sigma}{w}
\end{equation}

\noindent Therefore, for IID data, roughness linearly decreases with increased window size. Further, the kurtosis of random variable $S$, defined as the sum of independent random variables $R_1, R_2, ..., R_n$, is
\begin{equation}
\label{eq:kurtgeneral}
\mathrm{Kurt}[S] - 3 = \frac{1}{(\sum_{j=1}^n \sigma_j^2)^2}\sum_{i=1}^n\sigma_i^4(\mathrm{Kurt}[R_i] - 3)
\end{equation}
where $\sigma_i$ is the standard deviation of random variable $R_i$. In our case, $Y$ is the sum of $w$ IID random variables $X$~\cite{kurtosisIID}. Thus, Equation~\ref{eq:kurtgeneral} simplifies to
\begin{equation}
\label{eq:kurtiid}
\mathrm{Kurt}[Y] - 3 = \frac{\mathrm{Kurt}[X] - 3}{w}.
\end{equation}

Therefore, for IID series drawn from distributions with initial kurtosis less than 3, kurtosis monotonically increases with window length and for series drawn from distributions with initial kurtosis larger than 3, kurtosis monotonically decreases. 

In summary, these results indicate that for IID data, we can simply search for the largest window length that satisfies kurtosis constraint via binary search. Specifically, given a range of candidate window lengths, ASAP applies SMA with window length that is in the middle of the range. If the resulting smoothed series violates the kurtosis constraint, ASAP searches the smaller half of the range; otherwise, ASAP searches the large half. This binary search routine is justified because the roughness of the smoothed series monotonically decreases with window length (Equation~\ref{eq:roughnessiid}), and the kurtosis of the smoothed series monotonically decreases with window length or achieves its minimum at window length equals one (Equation~\ref{eq:kurtiid}).

However, many time series exhibit temporal correlations, which breaks the above IID assumption. This complicates the problem of window search, and we present a solution in the next subsection.

\subsection{\hspace{-0.2em}Optimization: Autocorrelation Pruning}
\label{asap:ac}
We have just shown that, for IID data, binary search is accurate, yet many time series are not IID;
instead, they are often periodic or exhibit other temporal correlations. For example, many servers and automated processes have regular workloads and exhibit periodic behavior across hourly, daily, or longer intervals.

To measure temporal correlations within a time series, we measure the time series \textit{autocorrelation}, or the similarity of a signal with itself as a function of the time lag between two points~\cite{timeseriesbook}. 
Formally, given a process $X$ whose mean $\mu$ and variance $\sigma^2$ are time independent (i.e., is a \textit{weakly stationary} process), denote $X_t$ as the value produced by a given run of the process at time t. The lag $\tau$ autocorrelation function ($\mathrm{ACF}$) on $X$ is defined as
\[ \mathrm{ACF}(X, \tau)  = \frac{\mathrm{cov}(X_t, X_{t+\tau})}{\sigma^2} =\frac{\mathbb{E}[(X_t - \mu)(X_{t+\tau} - \mu)]}{\sigma^2} \]
The value of the autocorrelation function ranges from [-1, 1], with 1 indicating perfect correlation, 0 indicating the lack of correlation and -1 indicating anti-correlation.

\subsubsection{Autocorrelation and Roughness}
As suggested above, we can take advantage of the periodicity in the original time series to prune the search space. Specifically, given the original time series $X: \{x_1, x_2, ..., x_N\}$, and the smoothed series $Y: \{y_1, y_2, ..., y_{N-w}\}$ obtained by applying a moving average of window length $w$, we show that
\begin{equation}
 \mathrm{roughness}(Y) = \frac{\sqrt{2}\sigma}{w}\sqrt{1 - \frac{N}{N-w}\mathrm{ACF}(X,w)}
 \label{eq:awesome}
\end{equation}
for a weakly stationary process $X$. We provide a full derivation of Equation~\ref{eq:awesome} in \cite{asaponline}, Appendix~A.1; however, intuitively, this equation illustrates that window length and autocorrelation both affect roughness. For example, consider a time series recording the number of taxi trips taken over 30-minute intervals. Due to the regularity of commuting routines, this time series exhibits autocorrelation across week-long periods (e.g., a typical Monday is likely to be much more similar to another Monday than a typical Saturday). Furthermore, a rolling weekly average of the number of trips should, in expectation, have a smaller variance than rolling 6-day averages: for example, if people are more likely to take taxis during weekdays than during weekends, then the average from Monday to Saturday should be larger than the average from Tuesday to Sunday. Therefore, window lengths that align with periods of high autocorrelation make the resulting series smoother. 

We experimentally validate this relationship on real world data (\cite{asaponline}, Appendix~A.1) and use this relationship to aggressively prune the space of windows to search (Section~\ref{sec:purning}).

\subsubsection{Autocorrelation and Kurtosis}
In addition to roughness, we also investigate the impact of temporal correlations on the kurtosis constraint. We start with an example that illustrates how choosing window lengths with high temporal correlation (i.e., autocorrelation) leads to high kurtosis.

Consider a time series (sparkline below, left) consisting of a sine wave with 640 data points. Each complete sine wave is 32 data points long, and in the region from 320th to 336th data point, the peak of the sine wave is taller than usual. When applying a window that are multiples of the period, the smoothed series (sparkline below, right) is zero everywhere except around the region where the peak is higher. The smoothed series in the latter case has higher kurtosis because it only contains one large deviation from the mean. In contrast, applying a moving average with window length that is not a multiple of the period will not highlight this peak.\\  
\includegraphics[width=0.45\columnwidth]{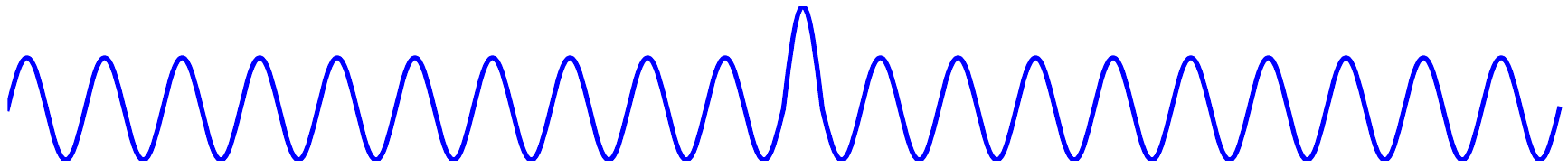} \qquad
\includegraphics[width=0.45\columnwidth]{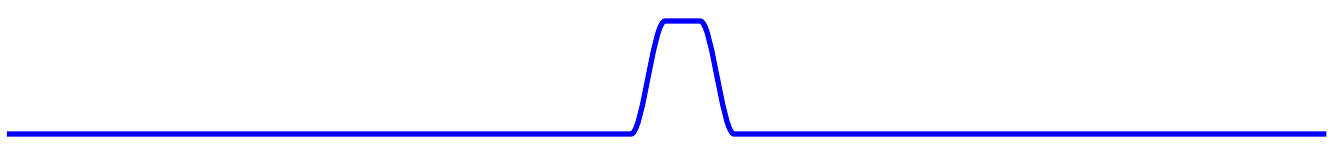}

This example illustrates the case when applying moving average with window lengths aligning with the period of the time series can not only remove periodic behavior from the visualization (therefore highlighting deviations from period to period), but also the kurtosis of the smoothed series is also larger at the periodic window size. In ASAP, we find that, empirically, if a candidate window that is aligned with the time series period does not satisfy the kurtosis constraint, it is rare that a nearby candidate window that is off the period would satisfy the constraint instead; moreover, such a nearby aperiodic window would likely result in a rougher series.

\begin{algorithm}[t!]
\footnotesize
\begin{algorithmic}
\State \textbf{Variables:} 
\State $\mathbf{X}$: time series; $\mathbf{candidates}$: array of candidate window lengths
\State $\mathbf{acf[w]}$: autocorrelation for $w$; $\mathbf{maxACF}$: maximum autocorrelation peak
\State $\mathbf{opt}$: a set of states for the current best candidate in the search, including\\ \quad\quad\{roughness, wLB, window, largestFeasibleIdx\}
\\\hrulefill
\Function{updateLB}{wLB, w} \Comment{Update lower bound}
\State \Return \textproc{max}(wLB, $w\sqrt{\frac{1 - maxACF}{1 - acf[w]}}$)
\EndFunction
\Function{isRougher}{currentBestWindow, w} \Comment{Compare roughness}
\State \Return $\frac{\sqrt{1 - acf[w]}}{w} > \frac{\sqrt{1 - acf[currentBestWindow]}}{currentBestWindow}$
\EndFunction
\Function{searchPeriodic}{X, candidates, opt}
\State N = candidates.length
\For{i $\in$ \{N, N-1, ..., 1\}} \Comment{Large to small}
\State w = candidates[i]
\If{ w $<$ opt.wLB} \Comment{Lower bound pruning}
\State break
\EndIf
\If {\textproc{isRougher}(opt.window, w)} \Comment{Roughness pruning}
\State continue
\EndIf
\State Y = \textproc{SMA}(X, w)
\If{  \textproc{roughness}(Y) $<$ opt.roughness \textbf{and} 
\State \textproc{kurt}(Y) $\geq$ \textproc{kurt}(X)} \Comment{Kurtosis constraint}
\State opt.window = w
\State opt.roughness = \textproc{roughness}(Y)
\State opt.wLB = \textproc{updateLB}(opt.wLB, w)
\State opt.largestFeasibleIdx = \textproc{max}(opt.largestFeasibleIdx, i)
\EndIf
\EndFor
\State \Return opt
\EndFunction
\end{algorithmic}
\caption{Search for periodic data}
\label{alg:periodic}
\end{algorithm}

\subsubsection{Pruning Strategies}
\label{sec:purning}
Following the above observations, ASAP adopts the following two pruning strategies. The corresponding pseudocode for ASAP's search is listed in Algorithm~\ref{alg:periodic}.

\minihead{Autocorrelation peaks} To quickly filter out suboptimal window lengths, ASAP searches for windows that correspond to periods of high autocorrelation. Specifically, ASAP only checks autocorrelation \textit{peaks}, which are local maximums in the autocorrelation function and correspond to periods in the time series. For periodic datasets, these peaks are usually much higher than neighboring points, meaning that the corresponding roughness of the smoothed time series is much lower. This is justified by Equation~\ref{eq:awesome}---all else equal, roughness decreases with the increase of autocorrelation.

Na\"ively computing autocorrelation via brute force requires $O(n^2)$ time; thus, a brute force this approach is unlikely to deliver speedups over the na\"ive exhaustive search for finding window length. However, we can improve the runtime of autocorrelation, to $O(n\log(n))$ time, using two Fast Fourier Transforms (FFT)~\cite{ACFviaFFT}. In addition to providing asymptotic speedups, this approach also allows us to make use of optimized FFT routines designed for signal processing, in the form of mature software libraries and increasingly common hardware implementations (e.g., DSP accelerators).

\minihead{Large to small} Since roughness decreases with window length (Equation~\ref{eq:awesome}, roughness is proportional to $\frac{1}{w}$), ASAP searches from larger to smaller window lengths. 
When two windows $w_1, w_2 (w_1 < w_2)$ have identical autocorrelation, the larger window will always have lower roughness under SMA. However, when the windows have different autocorrelations $a_1, a_2$, the smaller window $w_1$ will only provide lower roughness if $w_1 > w_2\sqrt{\frac{1-a_1}{1-a_2}}$. Moreover, since ASAP only considers autocorrelation peaks as candidate windows, $a_1$ is no larger than the largest autocorrelation peak in the time series, which we refer to as $maxACF$. Therefore, the smallest window $w_1$ that is able to produce smoother series than $w_2$ must satisfy
\begin{equation}
w_1  > w_2\sqrt{\frac{1-a_1}{1-a_2}} > w_2\sqrt{\frac{1-maxACF}{1-a_2}}
\label{eq:lb}
\end{equation}

If ASAP finds a feasible window length for smoothing relatively early in the search, it uses Equation~\ref{eq:lb} to prune smaller windows that will not produce a smoother series (\textproc{updateLB} in Algorithm~\ref{alg:periodic}). Similarly, once ASAP has a feasible window, it can also prune window candidates whose roughness estimate (via Equation~\ref{eq:awesome}) is larger than the current best (\textproc{isRougher} in Algorithm~\ref{alg:periodic}). In summary, the two pruning rules are complementary: the lower bound pruning reduces the search space from below, eliminating search candidates that are too small; the roughness estimate reduces the search space from above, further eliminating unpromising candidates above the lower bound. 

Our pruning strategies exploit temporal correlations, which will be less effective for aperiodic data. However, per our analysis in Section~\ref{sec:iid}, IID data is better-behaved under simple search. Therefore, ASAP falls back to binary search for aperiodic data. ASAP allows users to optionally specify a maximum window size to consider. Together, the search procedure is listed in Algorithm~\ref{alg:batch}.

\begin{algorithm}[t!]
\footnotesize
\begin{algorithmic}
\Function{findWindow}{X, opt}
\State candidates = \textproc{getAcfPeaks}(X)
\State opt = \textproc{searchPeriodic}(X, candidates, opt)
\State head = \textproc{max}(opt.wLB, candidates[opt.largestFeasibleIdx + 1])
\State tail = \textproc{min}(maxWindow, candidates[opt.largestFeasibleIdx + 1])
\State opt = \textproc{binarySearch}(X, head, tail, opt)
\State \Return opt.window
\EndFunction
\end{algorithmic}
\caption{Batch ASAP}
\label{alg:batch}
\end{algorithm}

\subsection{\hspace{-0.2em}Optimization: Pixel-aware Preaggregation}
\label{sec:pixel}
In addition to leveraging statistical properties of the data, ASAP can also leverage perceptual properties of the target devices. That is, ASAP's smoothed time series are designed to be displayed on devices such as computer monitors, smartphones, and tablet screens for human consumption.
Each of these target media has a limited resolution; as Table~\ref{tab:resolution} illustrates, even high-end displays such as the 2016 Apple iMac 5K are limited in horizontal resolution to 5120 pixels, while displays such as the 2016 Apple Watch contain as few as 272 pixels.
These pixel densities place restrictions on the amount of information that can be displayed in a plot.

ASAP is able to leverage these limited pixel densities to improve search time.
Specifically, ASAP avoids searching for window lengths that would result in more points than pixels supported by the target device.
For example, a datacenter server may report CPU utilization metrics every second (604,800 points per week). If an operator wants to view a plot of weekly CPU usage on her 2016 Retina MacBook Pro, she will only be able to see a maximum of 2304 distinct pixels as supported by the display resolution.
If ASAP smooths using a window smaller than 262 seconds (i.e., $\frac{604,800}{2304}$), the resulting plot will contain more points than pixels on the operator's screen (i.e., to display all information in the original time series, the slide size must be no larger than window length). As a result, this \textit{point-to-pixel} ratio places a lower bound on the window length that ASAP should search. In addition, the point-to-pixel ratio is also a useful proxy for the granularity of information content contained in a given pixel. While one could search for window lengths that correspond to sub-pixel boundaries, in practice, we have found that searching for windows that are integer multiples of the point-to-pixel ratio suffices to capture the majority of useful information in a plot. 
We provide an analysis in~\cite{asaponline} (Appendix~A.2), and empirically demonstrate these phenomena in Section~\ref{eval:opt}.

\begin{table}
\small
\center
\caption{Popular devices and search space reduction achieved via pixel-aware preaggregation for a series of 1M points.}
\vspace{-0.5em}
\begin{tabular}{ | l | l | r | }
  \hline			
  \textbf{Device} & \textbf{Resolution} & \textbf{Reduction on 1M pts} \\
  \hline		
 38mm Apple Watch &  272 x 340 & 3676x \hspace{24pt} \\	
 Samsung Galaxy S7 & 1440 x 2560 & 694x \hspace{24pt} \\ 
13'' MacBook Pro & 2304 x 1440 & 434x \hspace{24pt} \\
Dell 34 Curved Monitor & 3440 x 1440 & 291x \hspace{24pt}    \\
27'' iMac Retina & 5120 x 2880 & 195x \hspace*{24pt}\\
  \hline  
\end{tabular}
\label{tab:resolution}
\end{table}

Combined, these observations yield a powerful optimization for ASAP's search strategy. Given a target display resolution (or desired number of points for a plot), ASAP pushes this information into its search strategy by only searching windows that are integer multiples of point-to-pixel ratio. To implement this efficiently, ASAP preaggregates the data points according to groups of size corresponding to the point-to-pixel ratio, then proceeds to search over these preaggregated points. With this preaggregation, ASAP's performance is not dependent on the number of data points in the original time series but instead depends on the target resolution of the end device. As a result, in Section~\ref{sec:eval}, we evaluate ASAP's performance over different target resolutions and demonstrate scalability to millions of incoming data points per second. \vspace{-0.5em}

\subsection{\hspace{-0.2em}Streaming ASAP}
\label{sec:stream} 

ASAP is designed to process streams of time series and update plots as new data arrives. In this section, we describe how ASAP efficiently operates over data streams by combining techniques from traditional stream processing with constraints on human perception.

\minihead{Basic Operations} As new data points arrive, ASAP must update its smoothing parameters to accommodate changes in the trends, such as periodicity. As in Section~\ref{sec:pixel}, in the streaming setting, we can preaggregate data as it arrives according to the point-to-pixel ratio. However, as data transits the duration of time ASAP is configured to smooth (e.g., the last 30 minutes of readings), ASAP must remove outdated points from the window. To manage this intermediate state, ASAP adapts techniques from streaming processing that sub-aggregate input streams for performance gain. That is, sliding window aggregates such as SMA can be computed more efficiently by sub-aggregating the incoming data into disjoint segments (i.e., \textit{panes}) that are sizes of greatest common divisor of window and slide size~\cite{panes}. We can perform similar pixel-aware preaggregations for data streams using panes.

ASAP maintains a linked list of all subaggregations in the window and, when prompted, re-executes the search routine from the previous section. Instead of recomputing the smoothing window from scratch, ASAP records the result of the previous rendering request and uses it as a ``seed'' for the new search. Specifically, since streams often exhibit similar behavior over time, the previous smoothing parameter could possibly apply to the current request. In this case, ASAP starts the new search with a known feasible window length, which enables the roughness estimation pruning procedure (\textproc{isRougher} in Algorithm~\ref{alg:periodic}) to rule out candidates automatically.

\minihead{Optimization: On-demand updates}
\label{sec:ondemand}
A na\"ive strategy for updating ASAP's output is to update the plot upon arrival of each point. This is inefficient. For example, consider a data stream with a volume of one million points per second. Refreshing the plot for every data point requires updating the plot every 0.001 milliseconds. However, since humans can only perceive changes on the order of 60 events per second~\cite{60fps}, this update rate is unnecessary. With pixel-aware preaggregation, we would refresh for each aggregated data point instead,  the rate of which may still be higher than necessary. 
To visualize 10 minutes of data on a 27-inch iMac for example, pixel-aware preaggregation provides us aggregates data points that are 12ms apart (83Hz). As a result, we designed ASAP to only refresh at (configurable) timescales that are perceptible to humans. In our example above, a 1Hz update speed results in a 83$\times$ reduction in the number of calls to the ASAP search routine; this reduction means we will either use less processing power and/or be able to process data at higher volumes. In Section~\ref{eval:opt}, we empirically investigate the relationship between total runtime and refresh rate.

\minihead{Putting it all together}
\label{sec:together}
Algorithm~\ref{alg:stream} shows the full streaming ASAP algorithm. ASAP aggregates the incoming data points according to the desired point-to-pixel ratio, and maintains a linked list of the aggregates. After collecting a refresh-interval-time worth of aggregates, ASAP updates data points in the current visualization, and recalculates the autocorrelation (\textproc{updateACF}). ASAP then checks whether the window length from the last rendering request is still feasible (\textproc{checkLastWindow}). If so, ASAP uses this previous window length to quickly improve the lower bound for the new search. Otherwise, ASAP starts the new search from scratch. 

\begin{algorithm}[t!]
\footnotesize
\begin{algorithmic}
\State \textbf{Variables:}  $\mathbf{X}$: preaggregated time series; $\mathbf{interval}$: refresh interval\\
\hrulefill
\Function{checkLastWindow}{X, opt}
\State Y = \textproc{SMA}(X, opt.window)
\If{\textproc{kurt}(Y) $\geq$ \textproc{kurt}(X)}
\State update roughness and wLB for opt
\Else
\State re-initialize opt
\EndIf
\Return opt
\EndFunction
\Function{updateWindow}{X, interval}
\While {True}
\State collect new data points until interval
\State subaggregate new data points, and update X
\State \textproc{updateACF}(X)
\State opt = \textproc{checkLastWindow}(X, opt)
\State \textproc{findWindow}(X, opt)
\EndWhile
\EndFunction
\end{algorithmic}
\caption{Streaming ASAP}
\label{alg:stream}
\end{algorithm}

\section{Experimental Evaluation}
\label{sec:eval}

In this section, we experimentally evaluate the quality and efficiency of ASAP's visualizations via two user studies and a series of performance benchmarks. Our goal is to demonstrate that:
\begin{itemize}[noitemsep,parsep=.25em,topsep=.25em]
\item ASAP's visualizations improve both user accuracy and response time in identifying deviations (Section~\ref{eval:user-study}).
\item ASAP identifies high quality windows quickly (Section~\ref{eval:batchperf}). 
\item ASAP's optimizations---autocorrelation, pixel-aware aggregation and on-demand update---provide complementary speedups up to seven order-of-magnitude over baseline (Section~\ref{eval:opt}).
\end{itemize}

\subsection{\hspace{-0.2em}User Studies}
\label{eval:user-study}
We first evaluate the empirical effectiveness of ASAP's visualizations via two user studies. We demonstrate that ASAP visualizations lead to faster and more accurate identifications of anomalies.

\minihead{Visualization Techniques for Comparison} In each study, we
compare ASAP's visualizations to a set of alternatives
(cf. Section~\ref{sec:relatedwork}): $i)$ the original data, $ii)$ the
M4 algorithm~\cite{M4}, $iii)$ the Visvalingam-Whyatt algorithm~\cite{linesimp}, $iv)$ piecewise aggregate
approximation (PAA)~\cite{PAA} (PAA100 reduces the number of points to 100; PAA800 reduces to 800), and $v)$ an ``oversmoothed'' plot
generated by applying SMA with a window size of $\frac{1}{4}$ of the
number of points. All plots are rendered using an 800 pixel
resolution.

\minihead{Datasets} We select five publicly-available time series described in Table~\ref{tab:datasets} because each has known ground truth anomalies. We use this ground truth as a means of evaluating visualization quality by measuring users' ability to identify anomalous behaviors in the visualization and by assessing their preferences. Plots and text descriptions used in our user studies are available in Appendix~\ref{app:us} of the extended Technical Report~\cite{asaponline}.

\subsubsection{User Study I: Anomaly Identification}
\label{eval:detection-study}
\begin{figure}
\includegraphics[width=\linewidth]{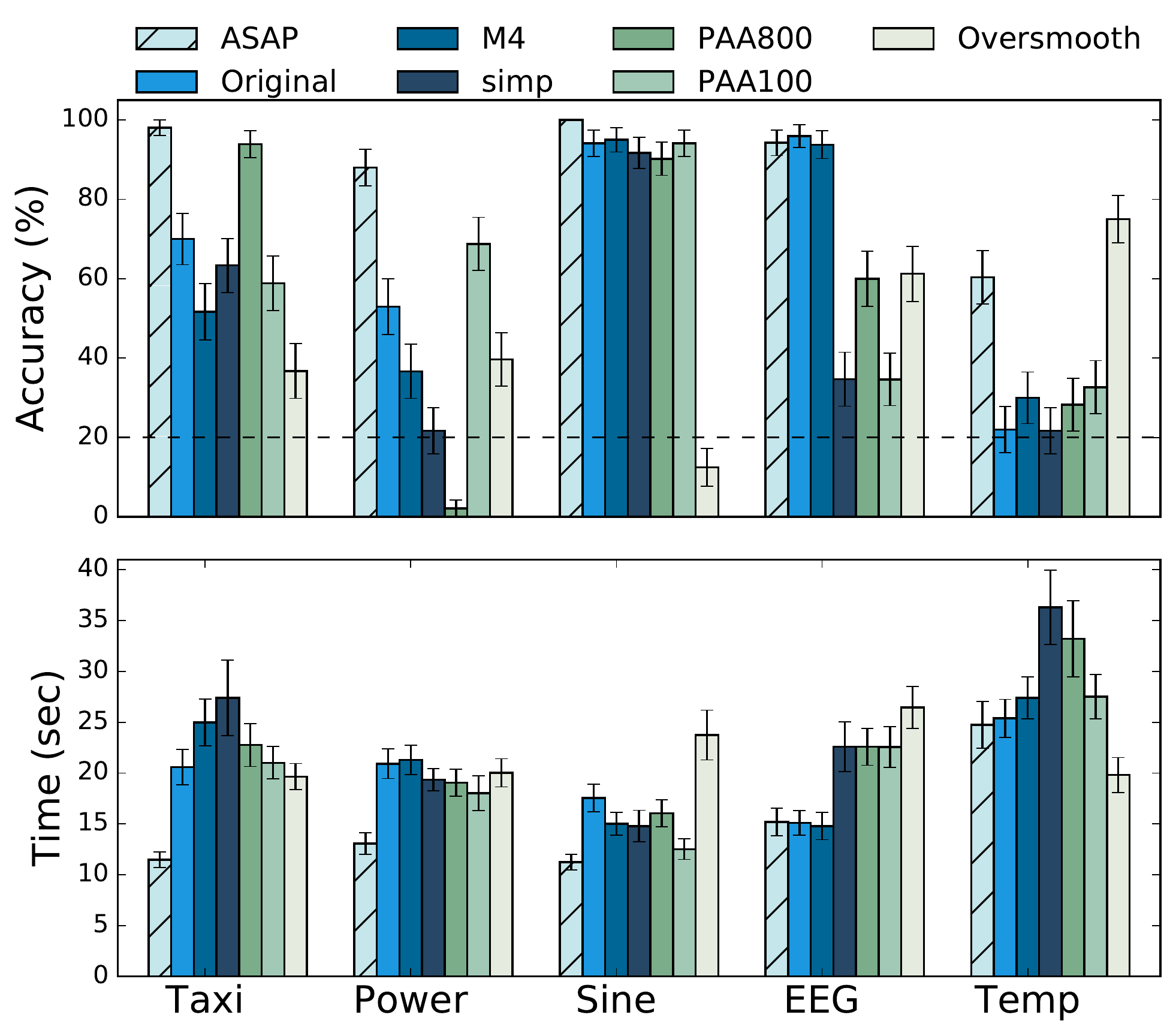}\vspace{-0.5em}
\caption{Accuracy in identifying anomalous regions and response times, with error bars indicating standard error of samples. On average, ASAP improves accuracy by 32.7\% while reducing response time by 28.8\% compared to other visualizations. }
\label{fig:us-region}
\end{figure}

To assess how different smoothing algorithms affect users' ability to identify anomalies in time series visualization, we ran a large-scale user study on Amazon Mechanical Turk, in accordance with Stanford University IRB guidelines.

In this first study, we presented users with textual descriptions of each dataset and anomaly, and asked them to select one out of the five equally-sized regions in a given time series visualization where the described anomaly occurred. Users performed anomaly identification using a single, randomly chosen visualization for each dataset, and, for each identification task, we recorded user's accuracy and response time. The user study involved 700 distinct Amazon Mechanical Turk workers, 
406 of whom self-reported as intermediate or expert users of Excel, 324 of whom self-reported as intermediate or expert users of databases, and 288 of whom self-reported seeing time series at least once per month. 

We report the accuracy and response time for the seven visualization techniques described above in Figure~\ref{fig:us-region}, where each bar in the plot represents an average of 50 users. When shown ASAP's visualizations, users were more likely to correctly identify the anomalous region, and to do so more quickly than alternatives. Specifically, users' accuracy of identifying the anomalous region increased by 21.3\% when presented with ASAP's visualizations instead of the original time series, and users did so 23.9\% more quickly. Compared to all other methods, users experience an average of 35.0\% (max: 43.1\%) increase in accuracy and 29.8\% (max: 33.8\%) decrease in response time with ASAP. ASAP led to most accurate results for all datasets except for the Temp dataset, in which the oversmooth strategy was able to better highlight (by 14.6\%) a large increasing temperature trend over several decades, corresponding to the rise of global warming~\cite{globalwarming}. However, ASAP results in 38.4\% more accurate identification than the raw data for this dataset. Overall, ASAP consistently produces high-quality plots, while the quality of alternative visualization methods varies widely across datasets. We provide additional results from a sensitivity analysis of the impact of roughness and kurtosis in~\cite{asaponline} (Appendix B.2), where we show that ASAP also outperforms alternative configurations in average accuracy and response time.

\subsubsection{User Study II: Visual Preferences}
\label{eval:preference-study}

In addition to the above user study, which was based on a large crowdsourced sample, we performed a targeted user study with 20 graduate students in Computer Science. We retained the same datasets and descriptions of dataset and anomaly from the previous study, and asked users to select the \emph{visualization} that best highlights the described anomaly in order to measure visualization preferences. In contrast with the previous study, due to smaller sample size, we presented a set of four visualizations--original, ASAP, PAA100, and oversmooth--anonymized and randomly permuted for each dataset.

\begin{figure}
\center
\includegraphics[width=\linewidth]{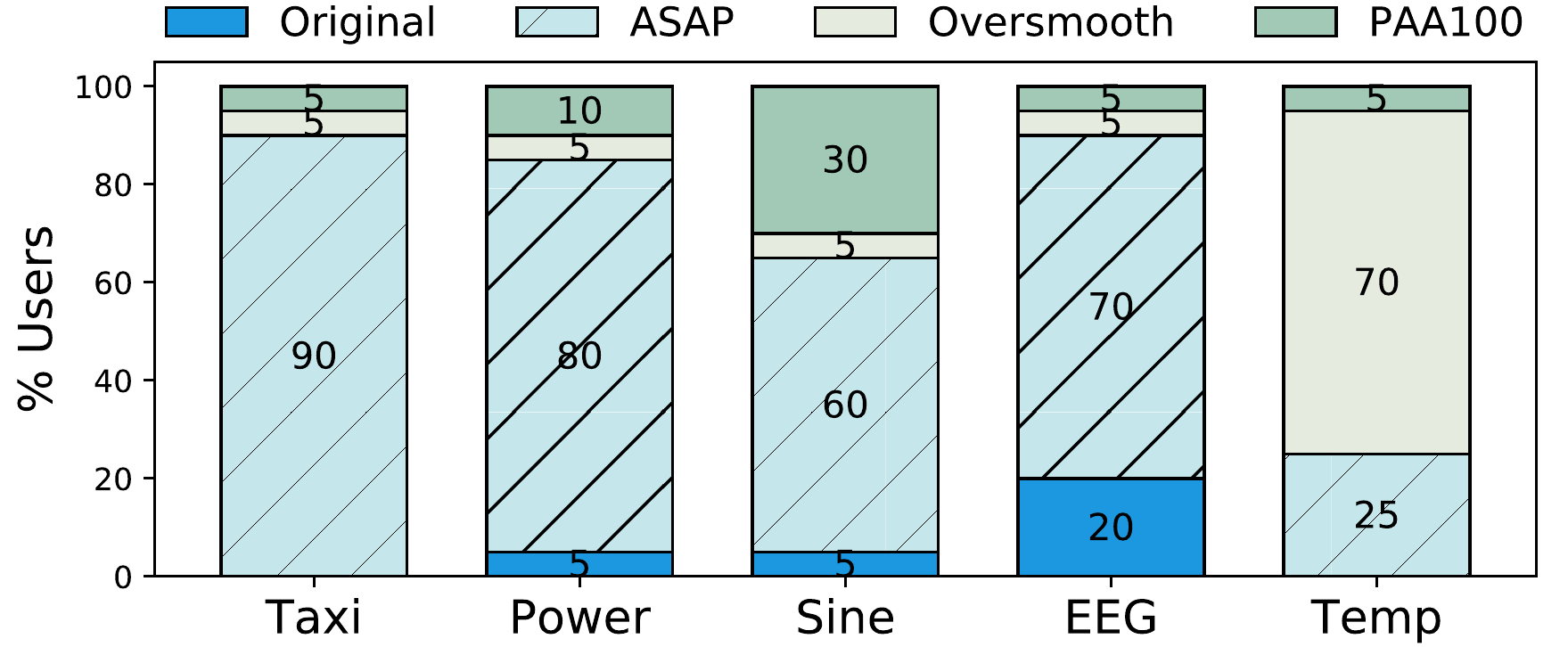}
\vspace{-1.5em}
\caption{Visual preference study. Users prefer ASAP 65\% of the time on average, and 59\% more often than the original time series.}
\label{fig:us-pref}
\end{figure}

Figure~\ref{fig:us-pref} presents results from this study. Across all five datasets, users preferred ASAP's visualizations as a means of visualizing anomalies in 65\% of the trials (random: 25\%). Specifically, for datasets Taxi (Figure~\ref{fig:us-taxi}, Appendix), EEG (Figure~\ref{fig:us-eeg}, Appendix), and Power (Figure~\ref{fig:us-power}, Appendix), over 70\% of users preferred ASAP's presentation of the time series. For these datasets, smoothing helps remove the high-frequency fluctuations in the original dataset and therefore better highlights the known anomalies. For dataset Sine (Figure~\ref{fig:us-sim}, Appendix), a simulated noisy sine wave with a small region where the period is halved, 60\% users chose ASAP, followed by 30\% choosing PAA100. In follow-up interviews, some users expressed uncertainty about this final plot: while the ASAP plot clearly highlights the anomaly, the PAA100 plot more closely resembles the description of the original signal. In the Temp dataset, 70\% of users chose the oversmoothed plot, and 25\% chose ASAP. For this dataset---which contains monthly temperature readings spanning over 250 years---aggressive smoothing better highlights the decade-long warming trend (Figure~\ref{fig:us-temp}, Appendix). In addition, no user preferred the original temperature plot, further confirming that smoothing is beneficial.

In summary, these results illustrate the utility of ASAP's target metrics in producing high-quality time series visualizations that highlight anomalous behavior.

\begin{table*}
\setlength\extrarowheight{1.1pt}
\scriptsize
\caption{Dataset descriptions and batch results from ASAP and exhaustive search over pre-aggregated data for target resolution 1200 pixels. ASAP finds the same choice of smoothing parameter as optimal, exhaustive search while searching an average of 13$\times$ fewer candidates. }\vspace{-1.5em}
\center
\begin{tabular}{ | p{2cm} | p{3.8cm} | p{1cm} | l || p{2.2cm}  | p{2.2cm} |}
  \hline
  \textbf{Dataset} & \textbf{Description} & \textbf{\# points} & \textbf{Duration} & \textbf{Exhaustive Search} & \textbf{ASAP}\\
\hline
gas\_sensor~\cite{UCI} & Recording of a chemical sensor exposed to a gas mixture & 4,208,261 & 12 hours & window size: 26 \newline \# candidates: 115 & window size: 26 \newline \# candidates: 7 \\
\hline
EEG~\cite{KeoghEEG} & Excerpt of
electrocardiogram & 45,000 & 180 sec & window size: 22 \newline \# candidates: 119 &  window size: 22 \newline \# candidates: 21 \\
\hline
Power~\cite{KeoghEEG} & Power consumption for
a Dutch research facility in 1997 & 35,040 & 35040 sec & window size: 16 \newline \# candidates: 115 & window size: 16 \newline \# candidates: 23 \\
\hline
traffic\_data~\cite{smartcity} & Vehicle traffic observed between two points for 4 months & 32,075 & 4 months  & window size: 84 \newline\# candidates: 120 & window size: 84  \newline\# candidates: 6 \\
\hline
machine\_temp~\cite{Numenta} & Temperature of an internal component of an industrial machine & 22,695 & 
70 days & window size: 44 \newline\# candidates: 125 & window size: 44  \newline \# candidates: 7\\
\hline
Twitter\_AAPL~\cite{Numenta} & 
A collection of Twitter mentions of Apple & 15,902 & 2 months & window size: 1 \newline\# candidates: 120 & window size: 1  \newline \# candidates: 7 \\
\hline  
ramp\_traffic~\cite{UCI} & Car count on a freeway ramp in Los Angeles & 8,640 & 1 month & window size: 96 \newline\# candidates: 117 & window size: 96  \newline\# candidates: 5 \\
\hline
sim\_daily~\cite{Numenta} & 
Simulated two week data with one abnormal day & 4,033 & 2 weeks & window size: 72 \newline \# candidates: 100 & window size: 72 \newline \# candidates: 5 \\
\hline			
Taxi~\cite{Numenta} & Number of NYC taxi passengers in 30 min bucket & 3,600 & 75 days & window size: 112\newline \# candidates: 120 & window size: 112  \newline\# candidates: 4\\
\hline
Temp~\cite{TSDL} & Monthly temperature in England from 1723 to 1970 & 2,976 & 248 years & window size: 112 \newline\# candidates: 120 & window size: 112 \newline \# candidates: 4\\
\hline
Sine~\cite{Keoghsin} & Noisy sine wave with an anomaly that is half the usual period & 800 & 800 sec & window size: 64 \newline\# candidates: 79 & window size: 64 \newline\# candidates: 6\\
\hline
\end{tabular}
\label{tab:datasets}
\end{table*}

\subsection{\hspace{-0.2em}Performance Analysis}
\label{eval:perf}
While the above user studies illustrate the utility of ASAP's
visualizations, it is critical that ASAP is able to render them
quickly and over changing time series. To assess
ASAP's end-to-end performance as well as the impact of each of its
optimizations, we performed a series of performance benchmarks.

\minihead{Implementation and Experimental Setup} We implemented an
ASAP prototype as an explanation operator for processing output data
streams in the MacroBase streaming analytics engine~\cite{MB}. We report results from evaluating the prototype on a
server with four Intel Xeon E5-4657L 2.4GHz CPUs containing 12 cores
per CPU and 1TB of RAM (although we use considerably less RAM in
processing). We exclude data loading time from our results but report
all other computation time.  We report results from the average of
three or more trials per experiment. We use a set of 11 of datasets of
varying sizes collected from a variety of application domains;
Table~\ref{tab:datasets} provides detailed descriptions of each
dataset; we provide plots from each experiment in~\cite{asaponline}.

\subsubsection{End-to-End Performance}
\label{eval:batchperf}

To demonstrate ASAP's ability to find high-quality window sizes quickly, we evaluate ASAP's window quality and search time compared to alternative search strategies. We compare to exhaustive search, grid search of varying step size (2, 10), and binary search.

First, as Table~\ref{tab:datasets} illustrates, with a target resolution of 1200 pixels, ASAP is able to find the same smoothing parameter as the exhaustive search for all datasets by checking an average of 8.64 candidates, instead of 113.64 candidates per dataset for the exhaustive search. For the Twitter\_AAPL dataset, both exhaustive search and ASAP leave the visualization unsmoothed; this time series (Figure~C.1, Appendix) is smooth except for a few unusual peaks, so further smoothing would have averaged out the peaks.
\begin{figure}
\includegraphics[width=\linewidth]{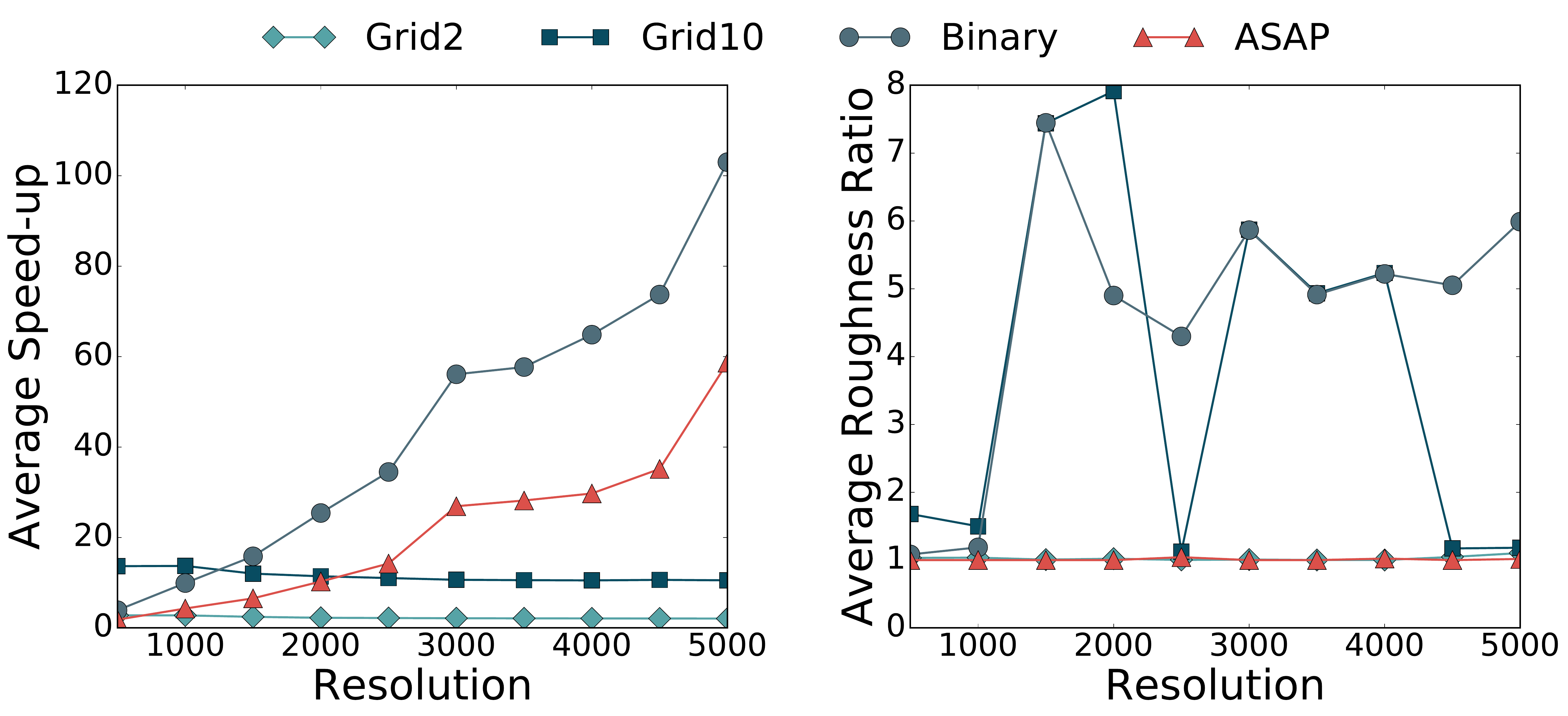}\vspace{-0.5em}
\caption{Throughput and quality of ASAP, grid search, and binary search over pre-aggregated time series according to varying target resolutions. Both plots report throughput and roughness compared to exhaustive search and report an average from the seven largest datasets in Table~\ref{tab:datasets}. ASAP exhibits similar speed-ups to binary search while retaining quality close to exhaustive search. ASAP's autocorrelation calculation incurs up to 50\% overhead compared to binary search but its results are up to 7.5$\times$ smoother.}
\label{fig:avg-batch}
\end{figure}

Second, we evaluate differences in wall-clock speed and achieved
smoothness. All algorithms run on preaggregated data, so the throughput difference is only caused by the difference in search strategies; we further investigate the impact of pixel-aware preaggregation in Section~\ref{eval:opt}. Figure~\ref{fig:avg-batch} shows that ASAP is able to achieve up to 60$\times$ faster search time than exhaustive search over pre-aggregated series,
with near-identical roughness ratio. ASAP's runtime performance scales
comparably to binary search, although it lags by up to 50\% due to its
autocorrelation calculation. However, while ASAP produces high-quality
smoothed visualizations, binary search is up to 7.5$\times$ rougher
than ASAP. Grid search with step size two delivers similar-quality
results as ASAP but fails to scale, while grid search with step size
ten delivers the worst overall results. In summary, end-to-end, ASAP
provides significant speedups over exhaustive search while retaining
its quality of visualization. We provide additional runtime comparison with PAA and M4 in~\cite{asaponline} (Appendix~A.3). 

\subsubsection{Impact of Optimizations}
\label{eval:opt}
In this section, we further evaluate
the contribution of each of ASAP's optimizations---autocorrelation
pruning, pixel-aware preaggregation, on-demand update---both individually
and combined.

\begin{figure}
\includegraphics[width=\linewidth]{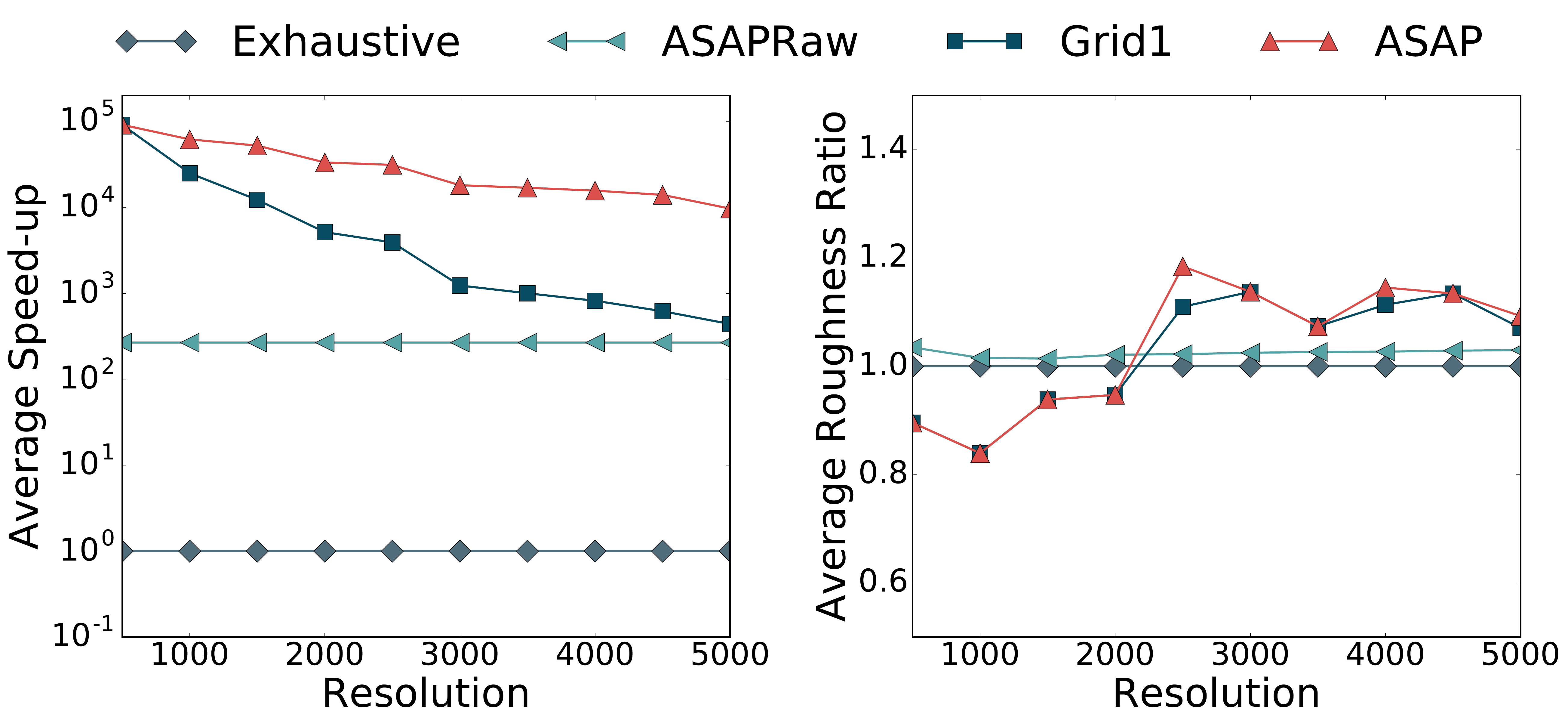}\vspace{-0.5em}
\caption{Throughput and quality of ASAP, exhaustive search on 
preaggregated time series over the baseline (exhaustive search over the original time series) under varying resolution. ASAP on aggregated time series is up to 4 orders of magnitude faster, while retaining roughness within $1.2\times$ the baseline. }
\label{fig:pixel-tpt}
\end{figure}

\minihead{Pixel-aware preaggregation}
We first perform a microbenchmark on the impact of pixel-aware preaggregation (Section~\ref{sec:pixel}) on both throughput and smoothness. Figure~\ref{fig:pixel-tpt} shows the throughput and quality of ASAP and exhaustive search with and without pixel-aware preaggregation under varying target resolutions. With pixel-aware pre-aggregation, ASAP achieves roughness within 20\% of exhaustive search over the raw series and sometimes outperforms exhaustive search because the initial pixel-aware preaggregation results in lower initial kurtosis. The preaggregation strategy enables a five and a 2.5 order-of-magnitude speedups over exhaustive search (Exhaustive) and ASAP on raw data (ASAPno-agg), respectively. In summary, pixel-aware preaggregation has a modest impact on result quality and massive impact on computational efficiency (i.e., sub-second versus hours to process 1M points). Should users desire exact result quality, they can still choose to disable pixel-aware preaggregation while retaining speedups from other optimizations. We provide additional analysis of pre-aggregation and additional experimental results in~\cite{asaponline} (Appendix~A.2). 

\begin{figure}
\center
\includegraphics[width=0.95\linewidth]{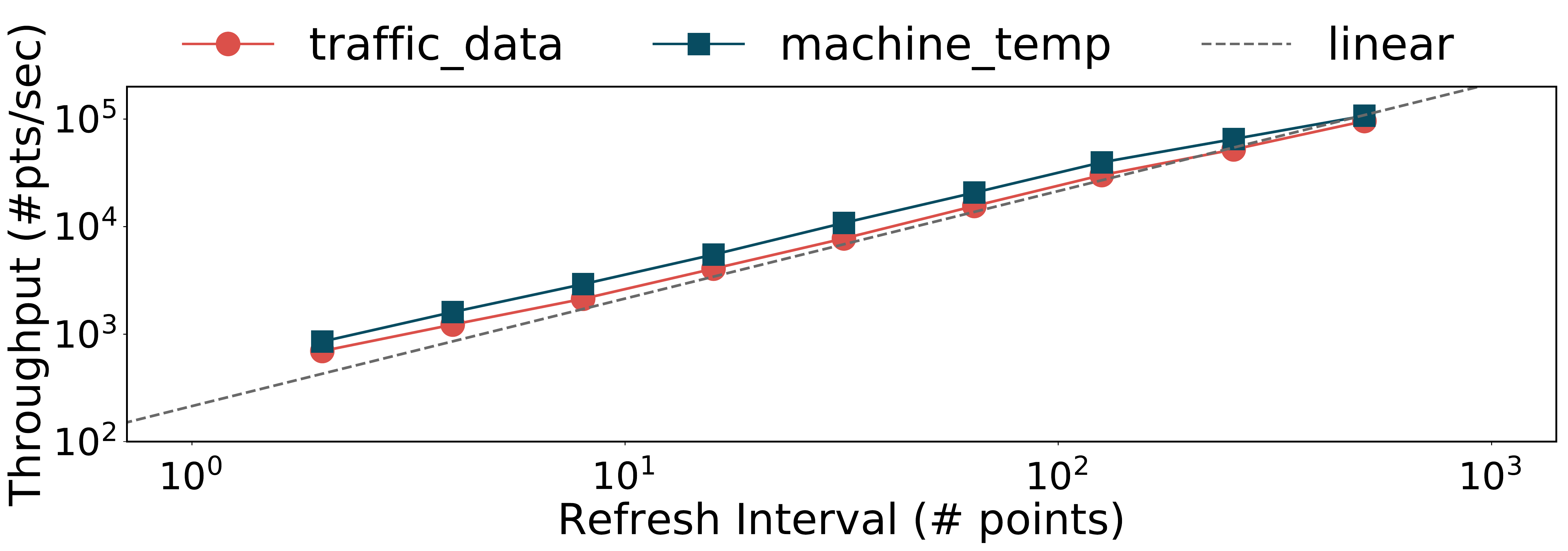}\vspace{-0.5em}
\caption{Throughput of streaming ASAP on two datasets, with varying refresh interval (measured in the number of points) for target resolution 2000 pixels in log-log scale. The plot captures a linear relationship between throughput and refresh interval as expected.}
\label{fig:stream}
\end{figure}

\minihead{On-demand update}
To investigate the impact of the update interval in the streaming setting (Section~\ref{sec:ondemand}), we vary ASAP's refresh rate and report throughput under each setting. The log-log plot (Figure~\ref{fig:stream}) shows a linear relationship between the refresh interval and throughput. This is expected because updating the plot twice as often means that it would take twice as long to process the same number of points. For fast-moving streams, this strategy can save substantial computational resources.

\begin{figure}
\center
\includegraphics[width=0.9\linewidth]{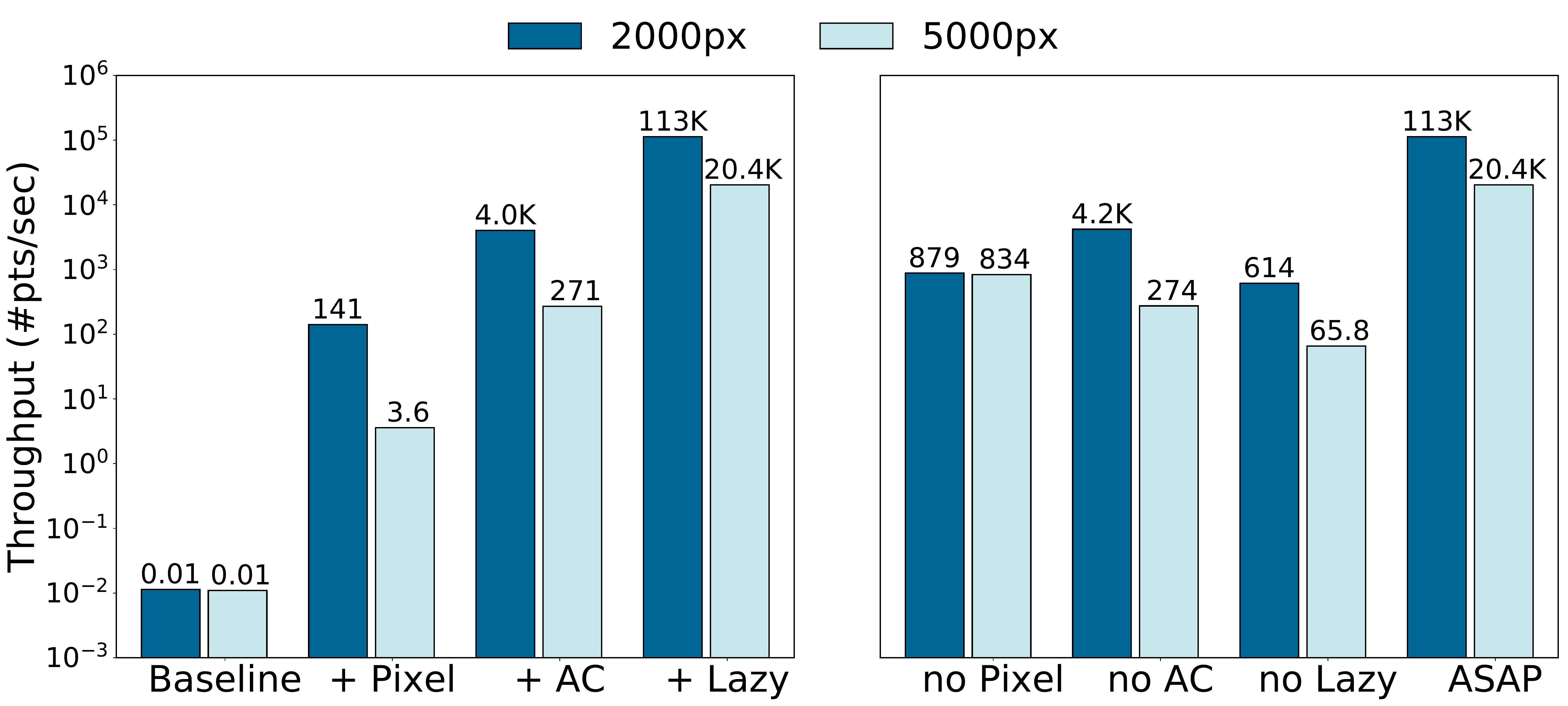}
 \caption{Factor analysis on machine\_temp dataset under two display settings. Cumulatively enabling optimizations shows that each contributes positively to final throughput; combined, the three optimizations enable seven orders of magnitude speedup over the baseline. In addition, removing each optimization decreases the throughput by two to three orders of magnitude.}
\label{fig:factor}
\end{figure}

\minihead{Factor Analysis} 
In addition to analyzing the impact of individual optimizations, we also investigate how ASAP's three main optimizations contribute to overall performance. Figure~\ref{fig:factor} (left) depicts a factor analysis, where we enable each optimization cumulatively in turn. Pixel-aware aggregation provides between two and four orders of magnitude improvement depending on the target resolution. Autocorrelation provides an additional two orders of magnitude. Finally, on demand update with a daily refresh interval (updating for every 288 points in the original series versus updating for each preaggregated point) provides another two order-of-magnitude speedups. These results demonstrate that ASAP's optimizations are additive and that end-to-end, optimized streaming ASAP is approximately seven orders of magnitude faster than the baseline.

To illustrate that no one ASAP optimization is responsible for all speedups, we perform a lesion study, where we remove each optimization from ASAP while keeping the others enabled (Figure~\ref{fig:factor}, right). Removing on-demand update, pixel-aware aggregation, and autocorrelation-enabled pruning each decreases the throughput by approximately two to three orders of magnitude, in line with results from the previous experiment. Without pixel-aware preaggregation, ASAP makes no distinction between higher and lower resolution setting, so the throughput for both resolutions are near-identical. In contrast, removing the other two optimizations degrades the performance for the higher resolution setting more. Thus, each of ASAP's optimizations is necessary to achieve maximum performance. 

\section{Related Work}
\label{sec:relatedwork}

ASAP's design draws upon work from several domains including stream
processing, data visualization, and signal processing.

\minihead{Time Series Visualization} 
Data visualization management systems that automate and recommend visualizations to users have recently become a topic of active interest in the database and human-computer interaction community~\cite{vizvision}. Recent systems including SeeDB~\cite{parameswaran2013seedb}, Voyager~\cite{wongsuphasawat2016voyager}, and ZenVisage~\cite{siddiqui2016zenvisage} focus on recommending visualizations for large-scale data sets, particularly for exploratory analysis of relational data~\cite{mackinlay2007show}. In this paper, we focus on the visualization of deviations within time series.

Within the time series literature, which spans simplification and reduction~\cite{PAA, RDP, linesimp, linesimpeval, financial, curve-segmentation}, information retrieval~\cite{hochheiser2004dynamic}, and data mining~\cite{nclusters,lin1995fast,lin2004visually,fu2011review}, visualization plays an important role in analyzing and understanding time series data~\cite{visualminingsurvey}. There are a number of existing approaches to time series visualization~\cite{aigner2011visualization}.
Perhaps most closely related is M4~\cite{M4}, which downsamples the original time series while preserving the shape---a perception-aware procedure~\cite{wunandi}. Unlike M4 and many existing time series visualization techniques, which focus on producing visually indistinguishable representations of the original time series (often using fewer points) by optimizing metrics such as pixel accuracy~\cite{RDP, linesimp, linesimpeval, financial, curve-segmentation}, ASAP visually highlights large-scale deviations in the time series by smoothing short term fluctuations.

To illustrate this difference in goals, we compared ASAP, M4 and the Visvalingam-Whyatt line simplification algorithm~\cite{linesimp} both on pixel accuracy (Appendix B.1,~\cite{asaponline}) and on end user accuracy of identifying anomalies (Section~\ref{eval:user-study}): ASAP is far worse at preserving pixel accuracy (up to 93\% worse, average: 91.8\% worse) than M4 but improves accuracy by up to 51\% (average: 26.7\%) for end-user anomaly identification tasks. These trends were similar for piecewise aggregate approximation~\cite{PAA}---which, in contrast, was \emph{not} originally designed for visualization. Despite differing objectives, we believe that pixel-preserving visualization techniques such as M4 are complementary to ASAP: a visualization dashboard could render the original time series using M4 and overlay with ASAP to highlight long-term deviations.

\minihead{Signal Processing} Noise reduction is a classic and extremely well studied problem in signal processing. Common reduction techniques include the wavelet transform~\cite{wavelet}, convolution with smoothing filters~\cite{SG,wiener}, and non-linear filters~\cite{bilateralfilter}. In this work, we study a specific type of linear smoothing filter---moving average---and the problem of its automatic parameter selection. Despite its simplicity, moving average is an effective time domain filter that is optimal at reducing random noise while retaining a sharp step response (i.e., rapid rise in the data)~\cite{Smith2003277}.

While there are many studies on parameter selection mechanisms for various smoothing functions~\cite{Marron1988}, the objective of most of the above selection criteria is to minimize variants of reconstruction error to the original signal. In contrast, ASAP's quality metric is designed to visually highlight trends and large deviations, leading to a different optimization strategy. Biomedical researchers have explored ideas of selecting a moving average window size that highlights significantly deviating region of DNA sequences~\cite{Tajima1991}. ASAP adopts a similar measure for quality---namely, preserving significant deviations in time series---but is empirically more efficient than the exhaustive approach described in the study.

\minihead{Stream Processing} To enable efficient execution, ASAP is architected as a streaming operator and adapts stream processing techniques. As such, ASAP is compatible with and draws inspiration from the rich existing systems literature on architectures for combining signal processing and stream processing functionality~\cite{girod2006wavescope, katsis2015combining}.

Specifically, aggregation over sliding windows has been widely recognized as a core operator over data streams. Sliding window semantics and efficient incremental maintenance techniques have been well-studied in the literature~\cite{Arasu:2004:RSC:1316689.1316720,Tangwongsan:2015:GIS:2752939.2752940}. ASAP adopts the sliding window aggregation model. However, instead of leaving users to select a window manually, in the parlance of machine learning, ASAP performs \textit{hyperparameter tuning}~\cite{hyperband} to automatically select a window that delivers smoothed plots that help improve end user's perception of long-term deviations in time series. We are unaware of any existing system---in production or in the literature---that performs this hyperparameter selection for smoothing time series plots. Thus, the primary challenge we address in this paper is efficiently and effectively performing this tuning via visualization-specific optimizations that leverage target display resolution, the periodicity of the signal, and on-demand updates informed by the limits of human perception.

\section{Conclusions}
\label{sec:conclusion}
In this paper, we introduced ASAP, a new data visualization operator that automatically smooths time series to reduce noise, prioritizing user attention towards systematic deviations in visualizations. We demonstrated that ASAP's target metrics---roughness and kurtosis---produce visualizations that enable users to quickly and accurately identify deviations in time series. We also introduced three optimizations---autocorrelation-based search, pixel-aware preaggregation and on-demand update---that provide order-of-magnitude speedups over alternatives without compromising quality. Looking forward, we are interested in further improving ASAP's scalability and in futher integrating ASAP with advanced analytics tasks including time series classification and alerting.

\section*{Acknowledgements}
We thank the many members of the Stanford InfoLab, the Monitorama
community, and Maneesh Agrawala for their valuable feedback on this work. This research was supported in part by affiliate members and other supporters of the Stanford DAWN project---Intel, Microsoft, Teradata, and VMware---as well as the Intel/NSF CPS Security grant \#1505728, the Secure Internet of Things Project, the Stanford Data Science Initiative, and industrial gifts and support from Toyota Research Institute, Juniper Networks, Visa, Keysight, Hitachi, Facebook, Northrop Grumman, NetApp, and Google.

\footnotesize
\bibliography{main} \bibliographystyle{abbrv}
\normalsize

\section*{APPENDIX}
\setcounter{section}{0}
 \setcounter{subsection}{0}
 \def\thesection{\Alph{section}}

\section{Additional Evaluations}
\counterwithin{figure}{section}
\setcounter{figure}{0}
\subsection{Roughness Estimate}
\label{app:roughness}
We first provide a full derivation for Equation~\ref{eq:awesome} in Section~\ref{asap:ac}. Given the original time series $X: \{x_1, x_2, ..., x_N\}$ (a weakly stationary process), and the smoothed series $Y: \{y_1, y_2, ..., y_{N-w}\}$ obtained by applying a moving average of window size $w$, we want to show that:
\begin{align*}
 \mathrm{roughness}(Y) = \frac{\sqrt{2}\sigma}{w}\sqrt{1 - \frac{N}{N-w}\mathrm{ACF}(X,w)}
\end{align*}
Note that in Equation~\ref{eq:roughness}, when the IID assumption does not hold,\\ $\mathrm{cov}(X_f, X_l) \neq 0$. 
The covariance of discrete two random variables $X, Y$ each with a set of $N$ equal-probability values is defined as:
\[\mathrm{cov}(X, Y) = \frac{1}{N} \sum_{i = 1}^N (x_i - \mathbb{E}(X))(y_i - \mathbb{E}(Y))\]
And for a discrete process, given N equi-spaced observations of the process $x_1, x_2, ..., x_N$, an estimate of the autocorrelation function at lag $k$ can be obtained by:
\[ \mathrm{ACF}(X, w) = \frac{\sum_{i = 1}^{N-w} (x_i - \bar{x})(x_{i + w} - \bar{x})}{\sum_{i = 1}^{N} (x_i - \bar{x})^2}\]
Therefore, we can rewrite the autocorrelation function as:
\[ \mathrm{ACF}(X, w) = \frac{(N-w)\mathrm{cov}(X_f, X_l)}{N\sigma^2}\text{,  or } \mathrm{cov}(X_f, X_l) = \frac{N\sigma^2}{N-w}\mathrm{ACF}(X, w)\]
Substituting $\mathrm{cov}(X_f, X_l)$ into (\ref{eq:roughness}), we obtain:
\begin{align*}
\mathrm{roughness}(Y) & = \frac{1}{w}\sqrt{\sigma^2 + \sigma^2 - 2\frac{N\sigma^2}{N-w}\mathrm{ACF}(X,w)}\\
& = \frac{\sqrt{2}\sigma}{w}\sqrt{1 - \frac{N}{N-w}\mathrm{ACF}(X,w)}
\end{align*}

\begin{figure}
\centering
\includegraphics[width=0.8\linewidth]{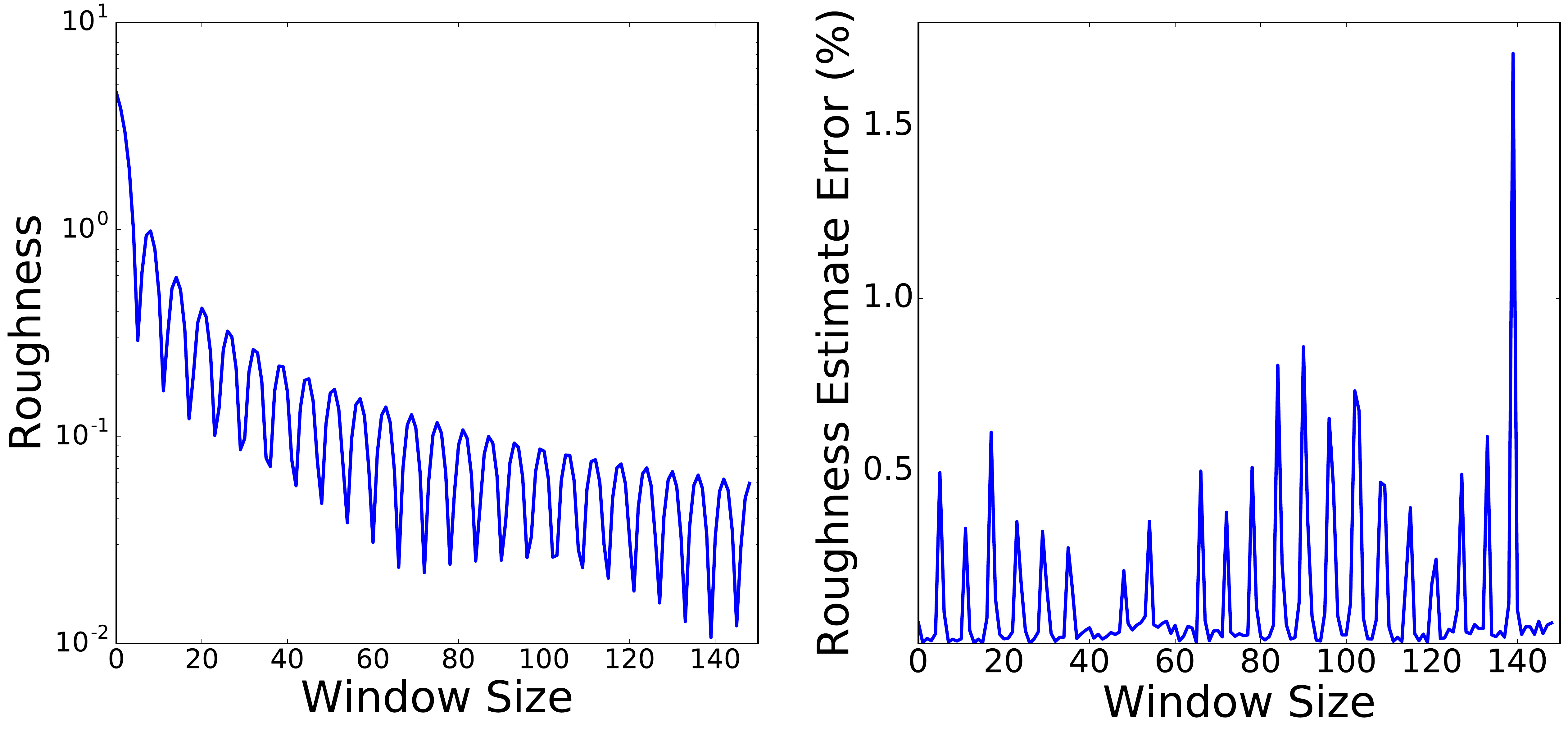} \hfill
\caption{True roughness and percent error of roughness estimation (Equation~\ref{eq:awesome}) over window sizes for dataset Temp. Estimate errors are with 1.2\% of the true value across all window sizes.}
\label{fig:roughnesserror}
\end{figure}

In addition, we empirically evaluate the accuracy of our roughness estimation (Equation~\ref{eq:awesome}) on the Temp dataset, and report the relative error in percent (Figure ~\ref{fig:roughnesserror}). For this time series, the roughness of the aggregated series drops sharply at window sizes around multiples of 6, which correspond to the autocorrelation peaks. Furthermore, estimated roughness (via Equation~\ref{eq:awesome}) is within 1.2\% of the true value across all window sizes.

\subsection{Pixel-aware Preaggregation} 
\label{app:PAA}
We first provide an analysis for the pixel-aware preaggregation
strategy. 
Given a time series of length $N$ sampled from uniform distribution and a target resolution of $t$ pixels, we have a point-to-pixel ratio of $p_a =\frac{N}{t}$. Let $w_{opt}$ be the window size that minimizes the roughness on the original time series. Note that searching on preaggregated data is equivalent to only selecting window sizes that are multiples of $p_a$. Since roughness decreases and kurtosis increases with window size, the optimal window size
over the preaggregated data is $w_{a} = \lfloor \frac{w_{opt}}{p_a}
\rfloor$, or $w_ap_a \leq w_{opt} < (w_a+1)p_a$. Therefore,
$\frac{w_{opt}}{w_ap_a} < \frac{(w_a + 1)p_a}{w_ap_a} =
\frac{w_a+1}{w_a}$. Recall roughness
scales proportionally with $\frac{1}{w}$ (Equation~\ref{eq:roughnessiid}), so preaggregation incurs a
penalty of no more than $\frac{w_a+1}{w_a}$ in roughness. Intuitively,
as optimal window size increases, quality of preaggregation increases and in
the limit, recovers the same solution as the search over the original data. We have a similar analysis for periodic data (roughness varies with $\frac{1}{w}\sqrt{1-ACF(w)}$). Let the autocorrelation corresponding to $w_{opt}$ be $ACF_{opt}$, and let 
the maximum change in autocorrelation along a window of size $p_a$ be $ACF_{\Delta}$. Specifically, while $w_{opt}$ may be able to pick an autocorrelation peak (i.e., a window size with high autocorrelation), searching on preaggregated data may
only come within $p_a$ of the peak. By examining the maximum
rise of the autocorrelation function over a period of length $p_a$, we
can bound the impact of roughness as above by
$\frac{w_a + 1}{w_a}\sqrt{\frac{1 - ACF_{opt} + ACF_{\Delta}}{1 - ACF_{opt}}}$. This implies that the impact of preaggregation on periodic data depends on the sharpness of
the autocorrelation function, which is in turn dataset-dependent. 
Our empirical results confirm that both of these effects are limited on real-world datasets.

\begin{figure}
\center
\includegraphics[width=0.75\linewidth]{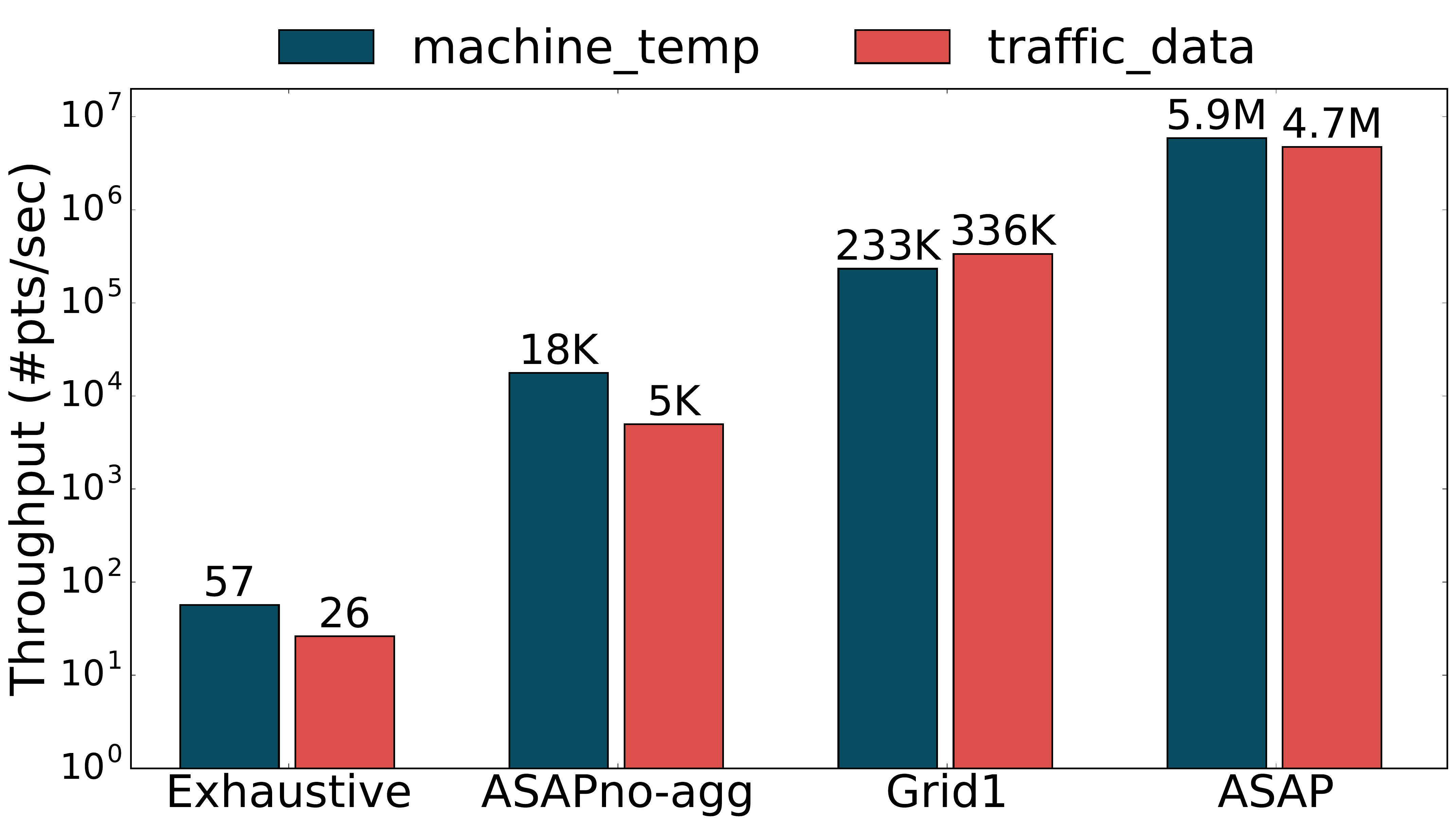}
\caption{Throughput of exhaustive search and ASAP on two datasets (machine\_temp, traffic\_data), without and with pixel-aware preaggregation for a target resolution of 1200 pixels. ASAP on preaggreaged data is up to 5 order of magnitude faster than exhaustive search on raw data (Exhaustive).}
\label{fig:pixel}
\end{figure}

We also provide additional performance evaluations for pixel-aware preaggregation. Figure~\ref{fig:pixel} shows the throughput of running exhaustive search and ASAP on two similar sized datasets (machine\_temp and traffic\_data), before and after applying the pixel-aware preaggregation. With a target resolution of 1200 pixels, ASAP on aggregated series is up to 5 order of magnitude faster compared to an exhaustive search on the original time series.

\begin{table}
\small
\center
\begin{tabular}{ | l | l |}
\hline
\textbf{Name} & \textbf{Description} \\
\hline
Exhaustive & Exhaustive search on raw time series\\
ASAPno-agg & ASAP on raw time series\\
Grid1 & Exhaustive search on preaggregated data\\
Grid2 & Exhaustive search with step size 2 on preaggregated data\\
Grid10 & Exhaustive search with step size 10 on preaggregated data\\
Binary & Binary search on preaggregated data\\
ASAP & ASAP on preaggregated data\\
\hline
\end{tabular}
\caption{Algorithms used in Section~\ref{eval:perf} }
\label{tab:algs}
\end{table}

\subsection{Runtime Comparison with O(n) Algorithms}
\begin{figure}\centering
\includegraphics[width=.9\linewidth]{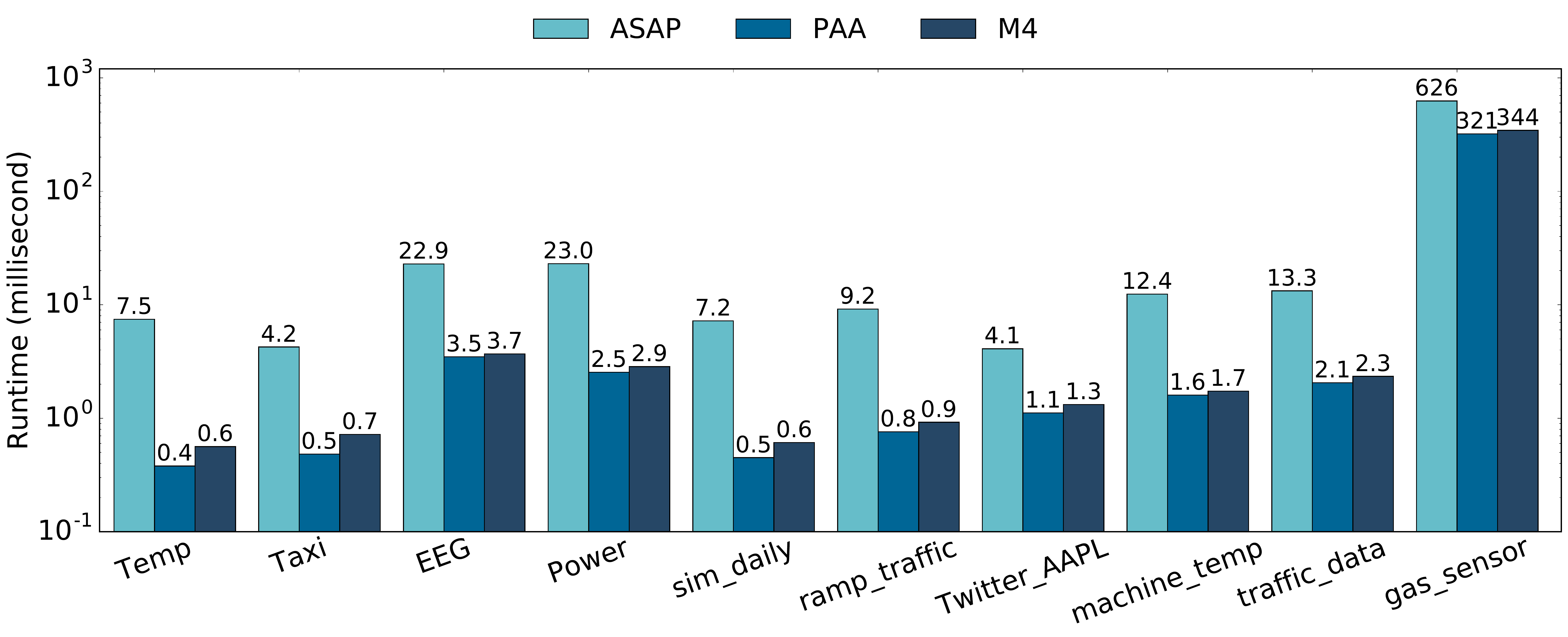}
\caption{Runtime of ASAP, PAA and M4 on datasets in Table 2.}
\label{fig:asap-paa}
\end{figure}

We report runtime for ASAP and two linear time algorithms PAA and M4 on ten datasets in Table 2 (target resolution 1200 pixels) in Figure~\ref{fig:asap-paa}. ASAP is up to 19.6 times slower than PAA and 13.2 times slower than M4. On average, ASAP, PAA and M4 complete in 72.9, 33.4 and 35.9 milliseconds respectively across all datasets.

\section{Additional User Study Results}
\label{app:us}
In this section, we present all time series plots and the accompanied text descriptions for the anomaly identification study (Section~\ref{eval:detection-study}). The visual preference study (Section~\ref{eval:preference-study}) uses a similar set of plots and text descriptions. 

\setcounter{figure} {2}
\begin{figure}\centering
\includegraphics[width=.75\linewidth]{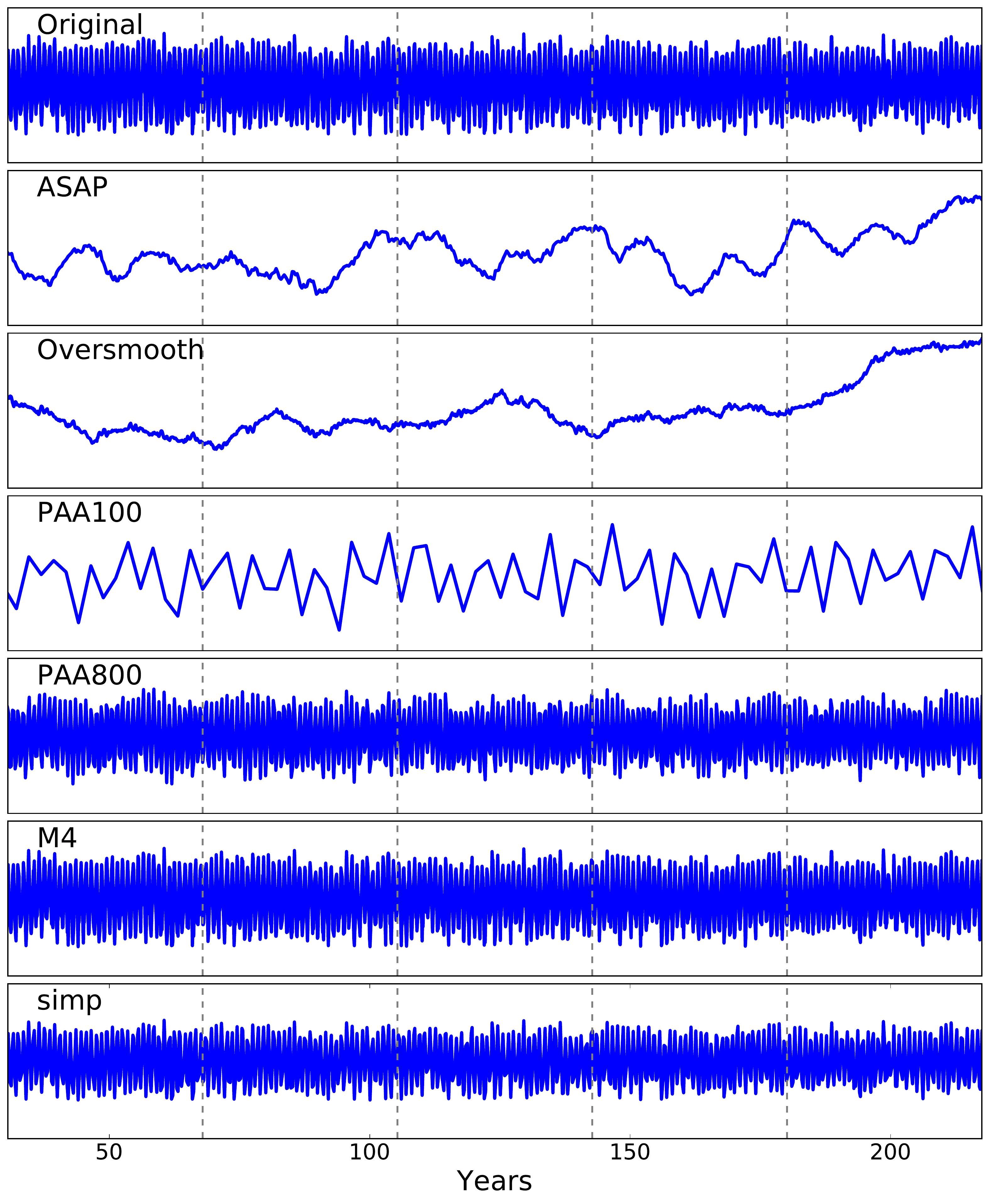}
\caption{User study plot for dataset Temp. We gave users the following description: ``The following plot depicts the temperature recorded in England in a 250-year period. The 1880s marked the end of a protracted period of cooling called the Little Ice Age, and the overall temperature started to increase afterwards. Which region of the following plot do you think this warming trend happened?"}
\label{fig:us-temp}
\end{figure}

\begin{figure}\centering
\includegraphics[width=.75\linewidth]{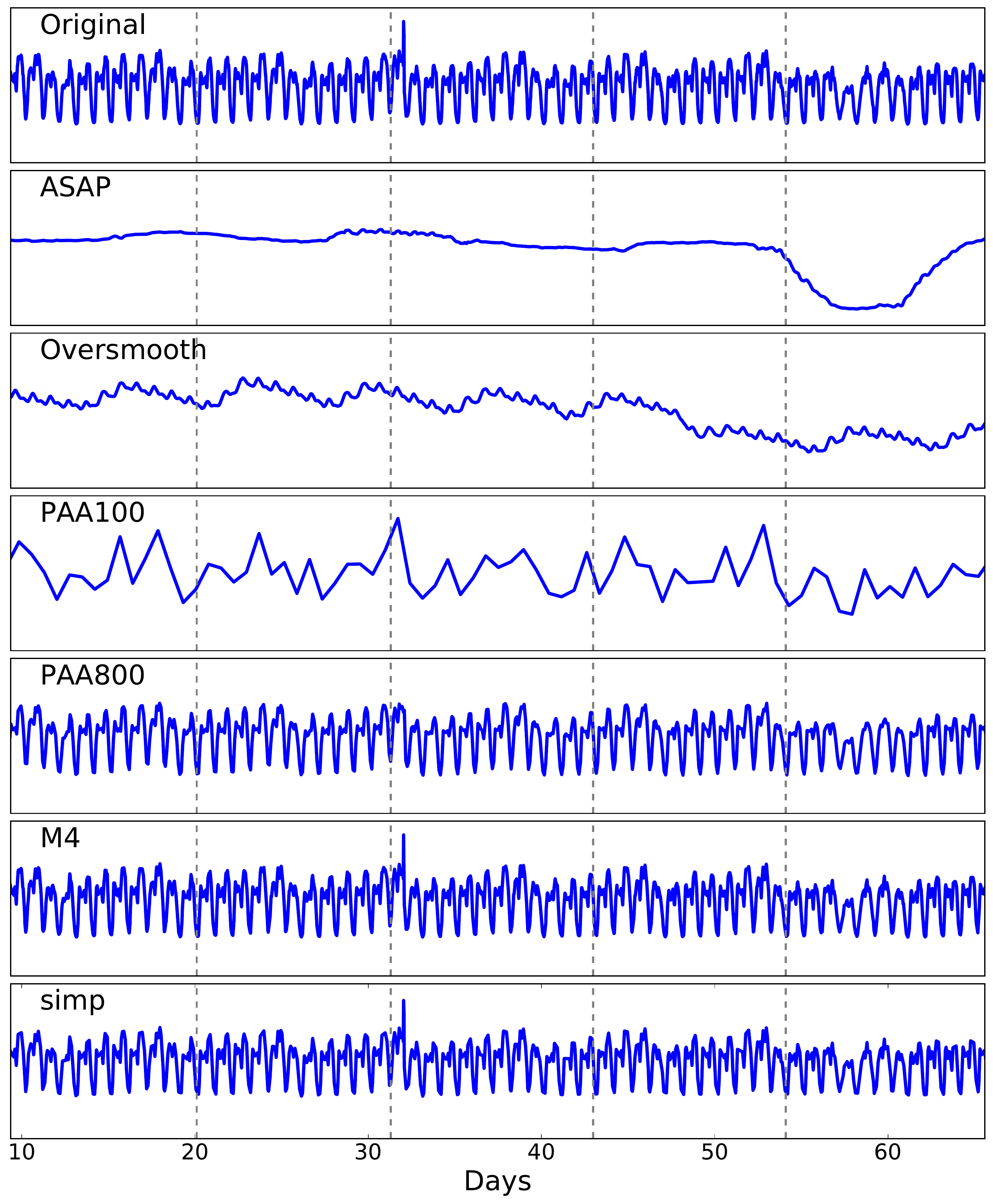}
\caption{User study plot for dataset Taxi. We gave users the following description: ``The following plot depicts the volume of taxicab trips in New York City in a 2 month period in 2014. The volume of taxicab trips dropped sustainedly during the week of November 24th (due to Thanksgiving). Which region of the following plot do you think this sustained drop in volume happened?"}
\label{fig:us-taxi}
\end{figure}

\begin{figure}\centering
\includegraphics[width=.75\linewidth]{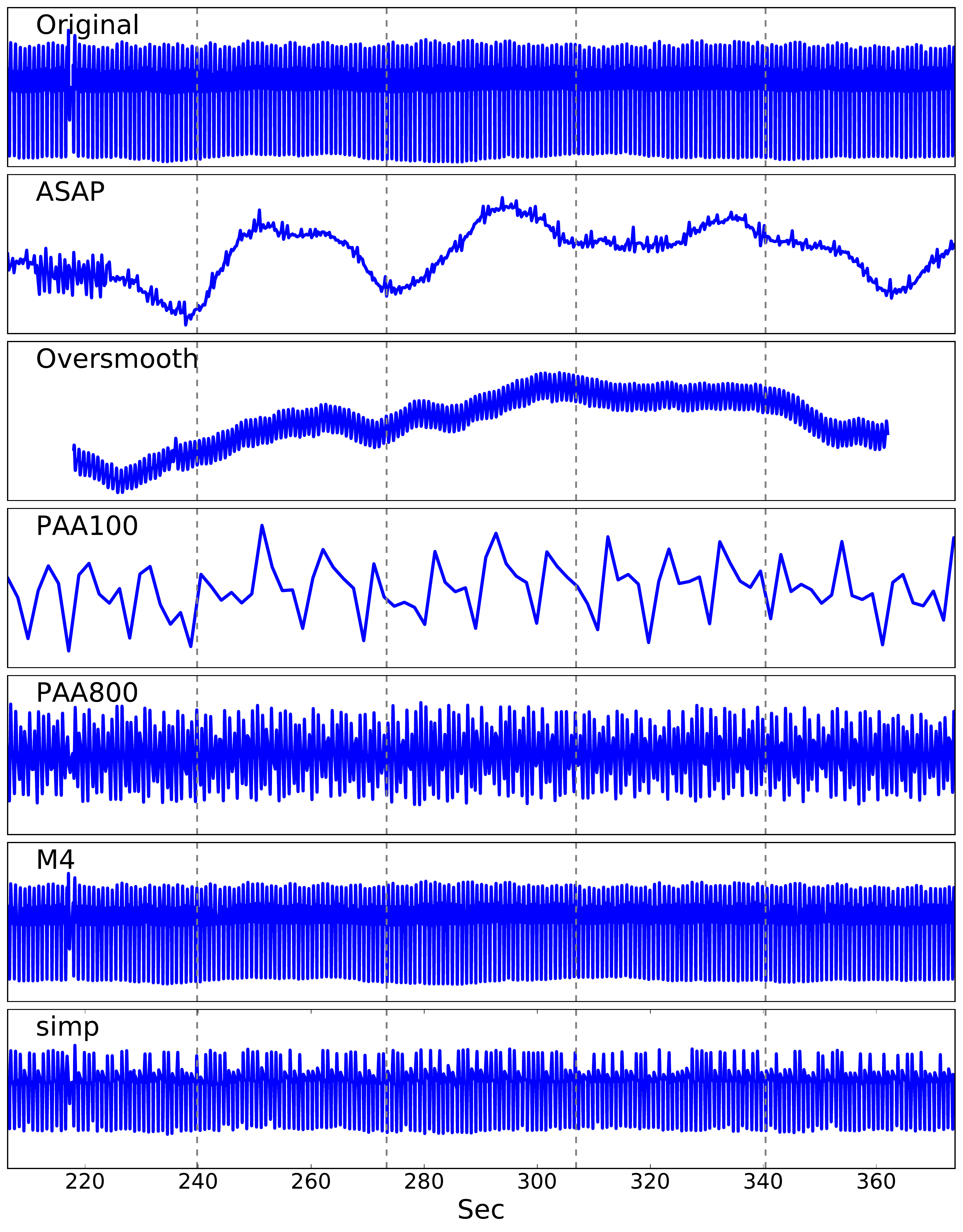}
\caption{User study plot for dataset EEG. We gave users the following description: ``The following plot depicts 22,500 readings measuring a patient's brainwaves (EEG activity). The EEG segment shown below contains an abnormal pattern (corresponding to a premature ventricular contraction).  Which region of the following plot do you think this abnormal pattern occurred?"}
\label{fig:us-eeg}
\end{figure}

\begin{figure}\centering
\includegraphics[width=.75\linewidth]{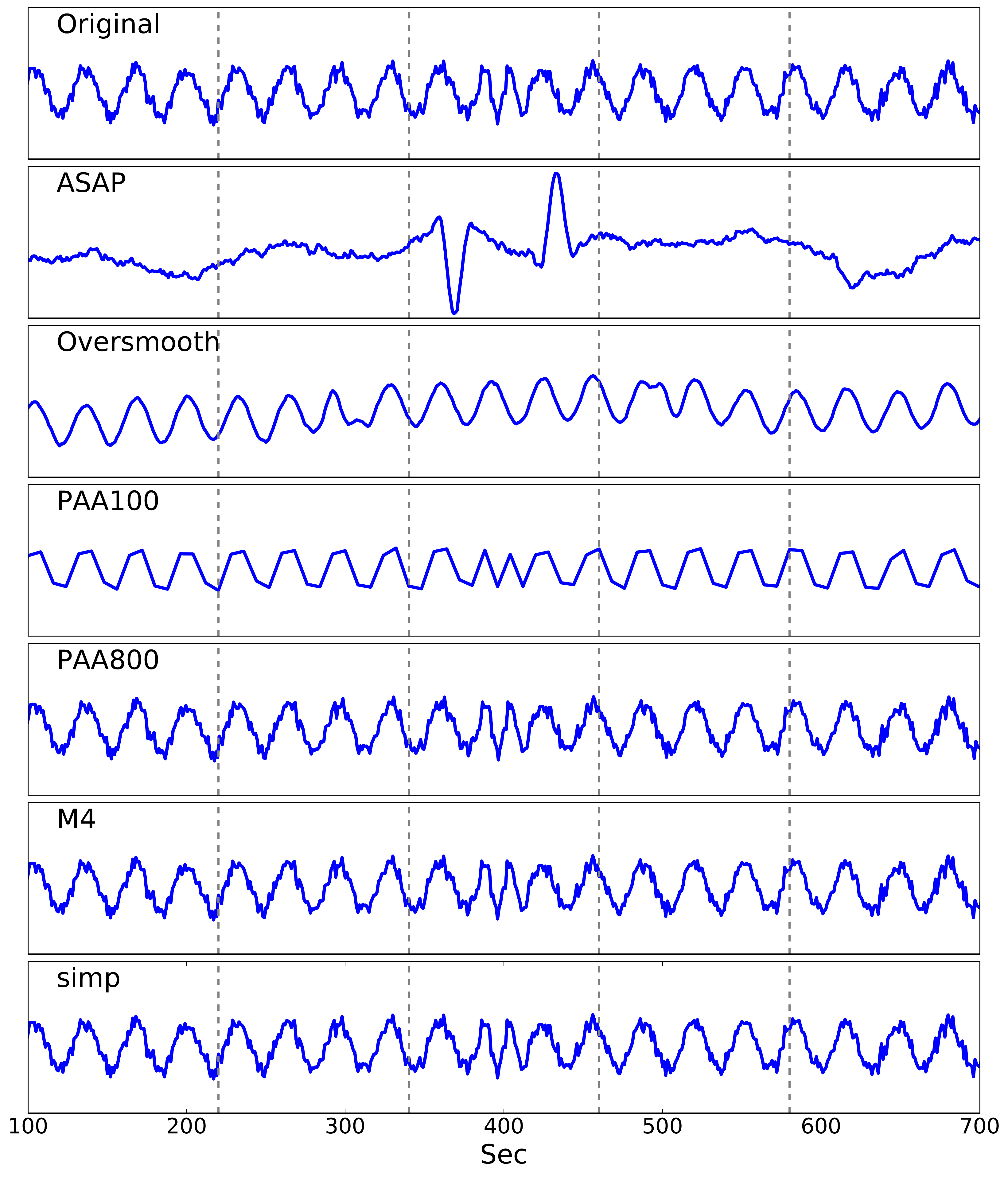}
\caption{User study plot for dataset Sine. We gave users the following description: ``The following plot depicts 800 readings from a time varying signal. At some point, the signal experienced an unusual deviation from its regular behavior. Which region of the following plot do you think this deviation happened?"}
\label{fig:us-sim}
\end{figure}

\begin{figure}\centering
\includegraphics[width=.75\linewidth]{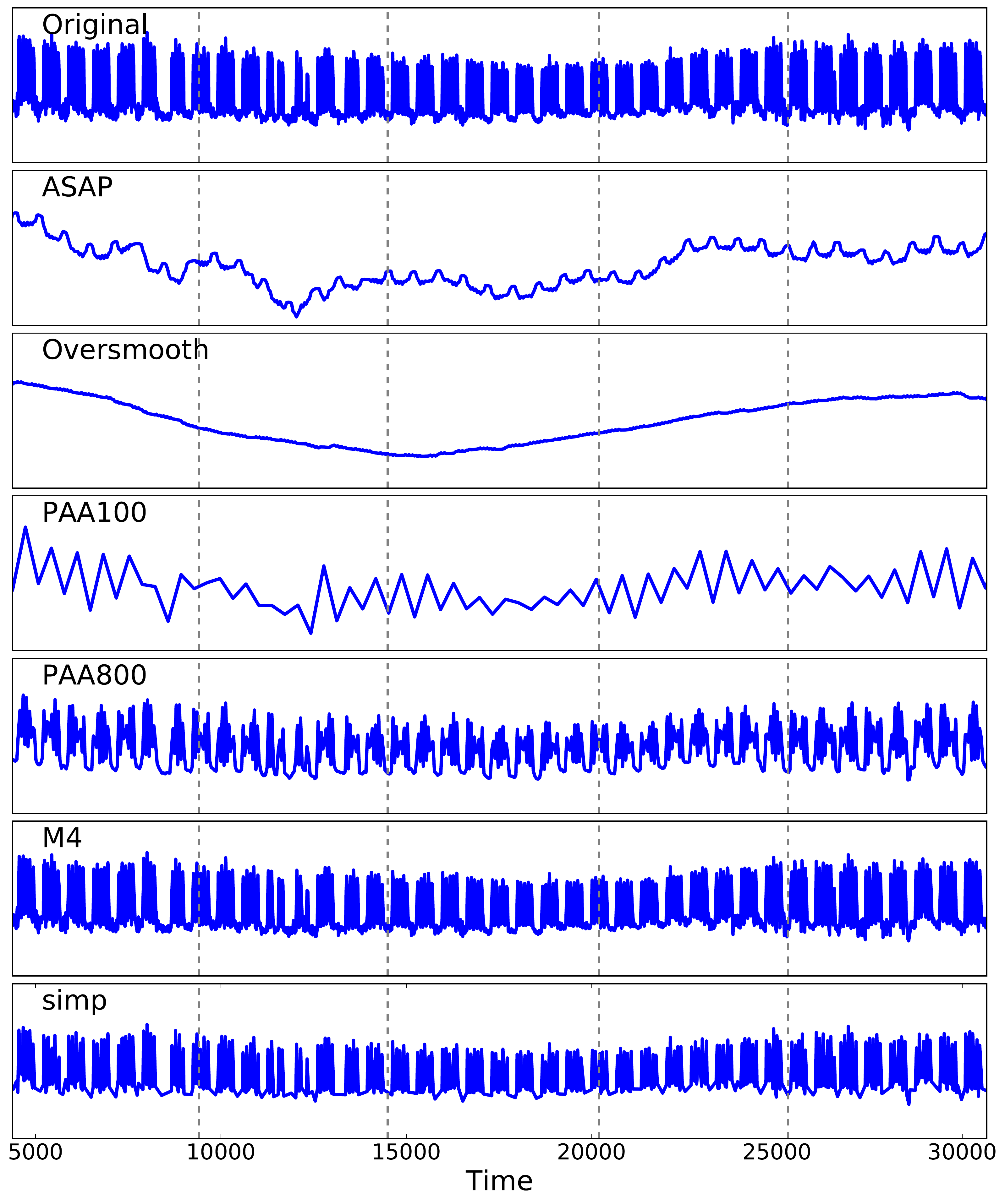}
\caption{User study plot for dataset Power. We gave users the following description: ``The following plot depicts one year of power demand at a Dutch research facility. The power demand temporarily dips during the Ascension Thursday holiday.  Which region of the following plot do you think this dip in power demand occurred?"}
\label{fig:us-power}
\end{figure}

\subsection{Pixel Error}
We first report pixel errors for all data reduction algorithms compared in the anomaly identification study (Table~\ref{tab:pixelerror}). Consistent with findings of previous studies, moving average based algorithms and line simplification algorithms result in larger pixel errors compared to MinMax based algorithm such as M4~\cite{M4}. However, to reiterate on the difference in goals, ASAP is designed to minimize roughness rather than pixel errors. As we demonstrated in the user study (Section 5.1.1), ASAP is able to better highlight long-term deviations in the time series despite its large pixel errors.

\begin{table}
\small
\center
\begin{tabular}{ | l | l | l | l | l |}
\hline
 \textbf{Dataset} & \textbf{ASAP} & \textbf{M4} & \textbf{Line Simplification} & \textbf{PAA800} \\
\hline
Temp & 0.94 & 0.02 & 0.06 & 0.36 \\
Taxi & 0.94 & 0.02 & 0.05 & 0.22 \\
EEG & 0.92 & 0.02 & 0.21 & 0.61 \\
Sine & 0.93 & 0 & 0 & 0 \\
Power & 0.94 & 0.04 & 0.17 & 0.56 \\
\hline
\end{tabular}
\caption{Pixel error of ASAP, M4, Visvalingam-Whyatts line simplification algorithm, and PAA800 on the user study datasets. }
\label{tab:pixelerror}
\end{table}

\subsection{Sensitivity Study}
\minihead{Roughness} In addition to the anomaly identification study presented in Section~\ref{eval:detection-study}, we varied the target roughness for each dataset and measured the impact on end user's accuracy and response time. Specifically, we used the roughness of the ASAP plot for each dataset as the reference, and generated plots that have 8 times (8x), 4 times (4x), 2 times (2x) and half (1/2x) the roughness. We report results in Figure~\ref{fig:kurt-smooth}, where bar represents the average accuracy and response time from about 50 Amazon mechanical turk workers. While the results varies across each dataset, we observe that less smooth plots result in lower average accuracy (61.5\% for 8x and 55.8\% for 4x) compared to smoother plots (78.6\% for 2x and 79.8\% for 1/2x). Overall, ASAP achieves the highest average accuracy and the lowest response time among all configurations.
\setcounter {figure} {0}
\begin{figure}\centering
\includegraphics[width=\linewidth]{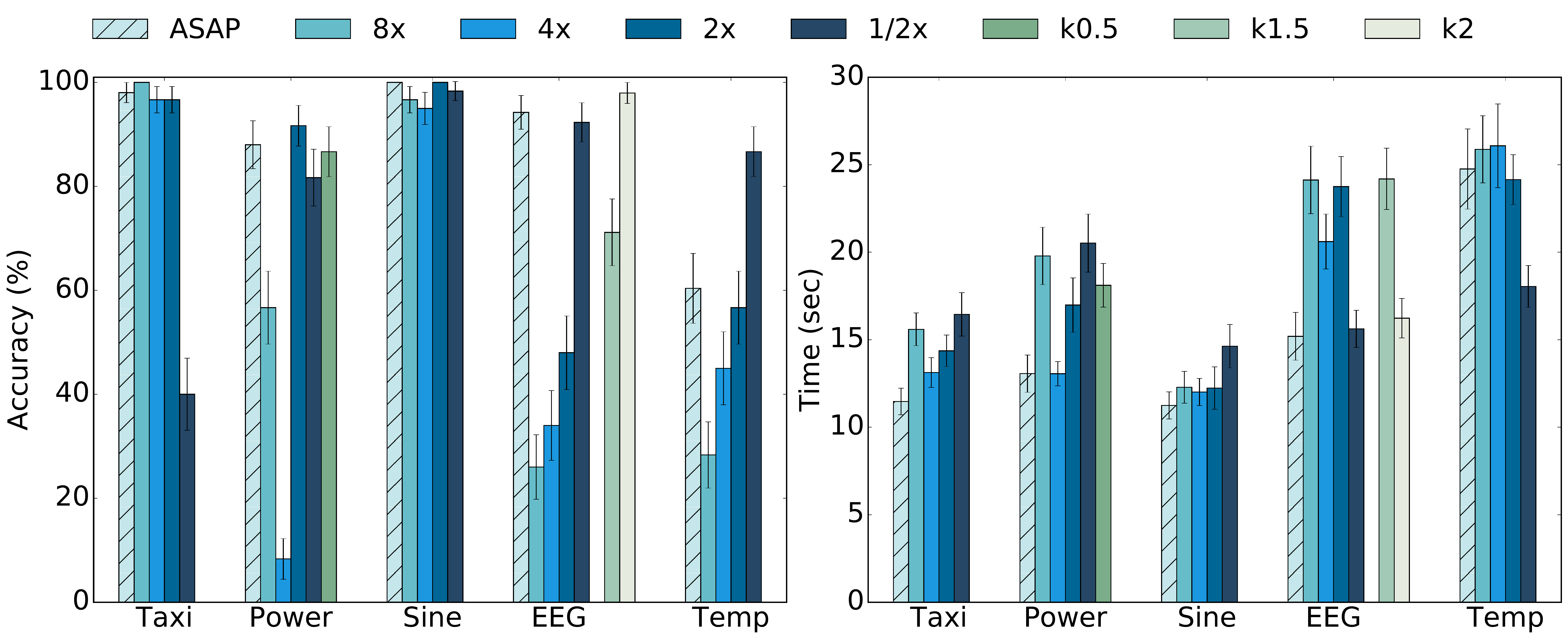}
\caption{Impact of roughness and kurtosis on user's accuracy and response time for identifying anomaly in the time series. }
\label{fig:kurt-smooth}
\end{figure}

\minihead{Kurtosis} Similar to the roughness study, we varied the kurtosis constraint to preserving 0.5x (k0.5), 1.5x (k1.5) and 2x (k2) of kurtosis of the original time series. For three out of the five datasets, varying the kurtosis constraint didn't affect the final visualization, since the ASAP visualization already has a relatively small roughness and high kurtosis. We report additional kurtosis results for the Power and EEG dataset in Figure~\ref{fig:kurt-smooth}. Overall, we find that roughness has a larger impact on end results compared to kurtosis.

\begin{figure}\centering
\begin{subfigure}[b]{0.9\columnwidth}
  \includegraphics[width=\textwidth]{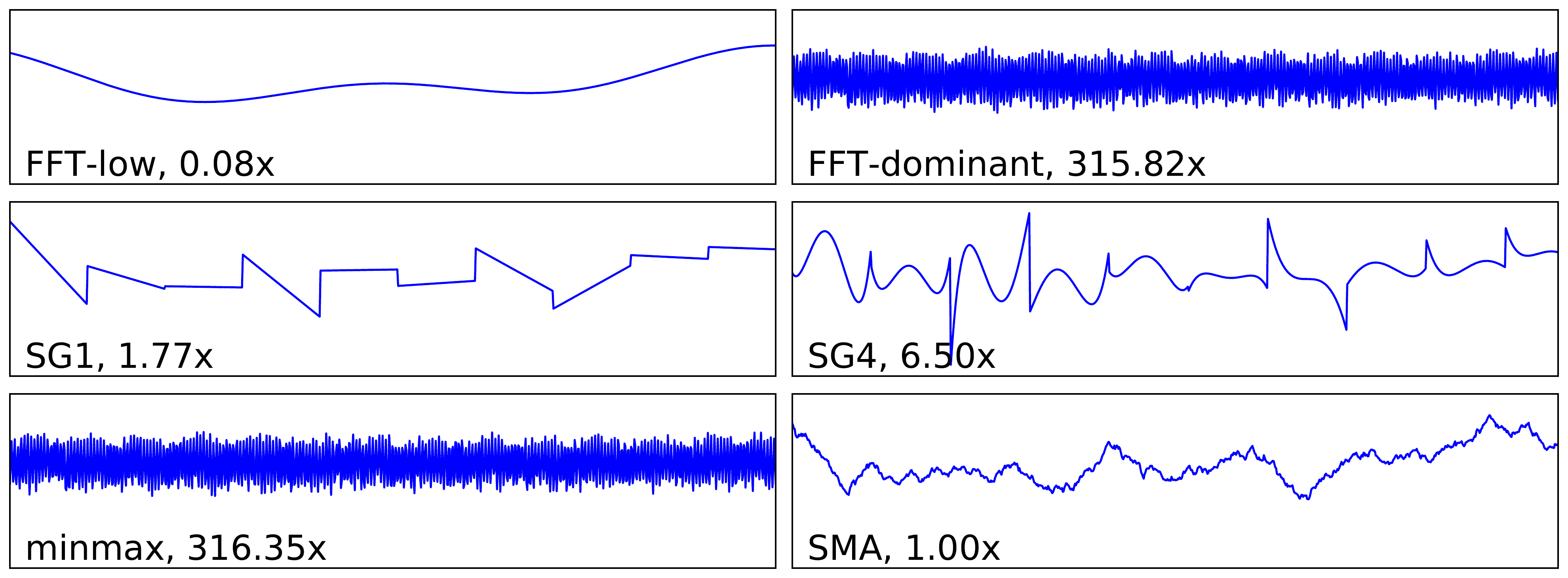}
  \caption{Temp}
  \label{fig:func_14}
\end{subfigure}
\begin{subfigure}[b]{0.9\columnwidth}
  \includegraphics[width=\textwidth]{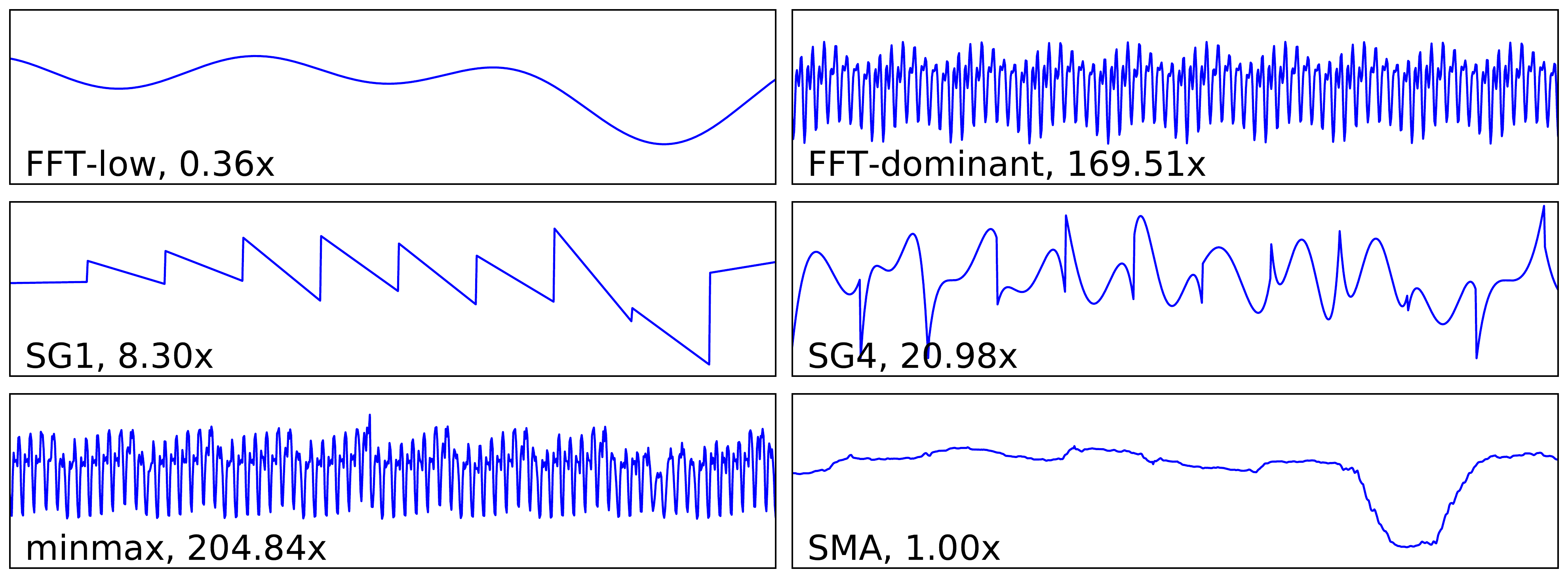}
  \caption{Taxi}
  \label{fig:func_27}
\end{subfigure}
\begin{subfigure}[b]{0.9\columnwidth}
  \includegraphics[width=\textwidth]{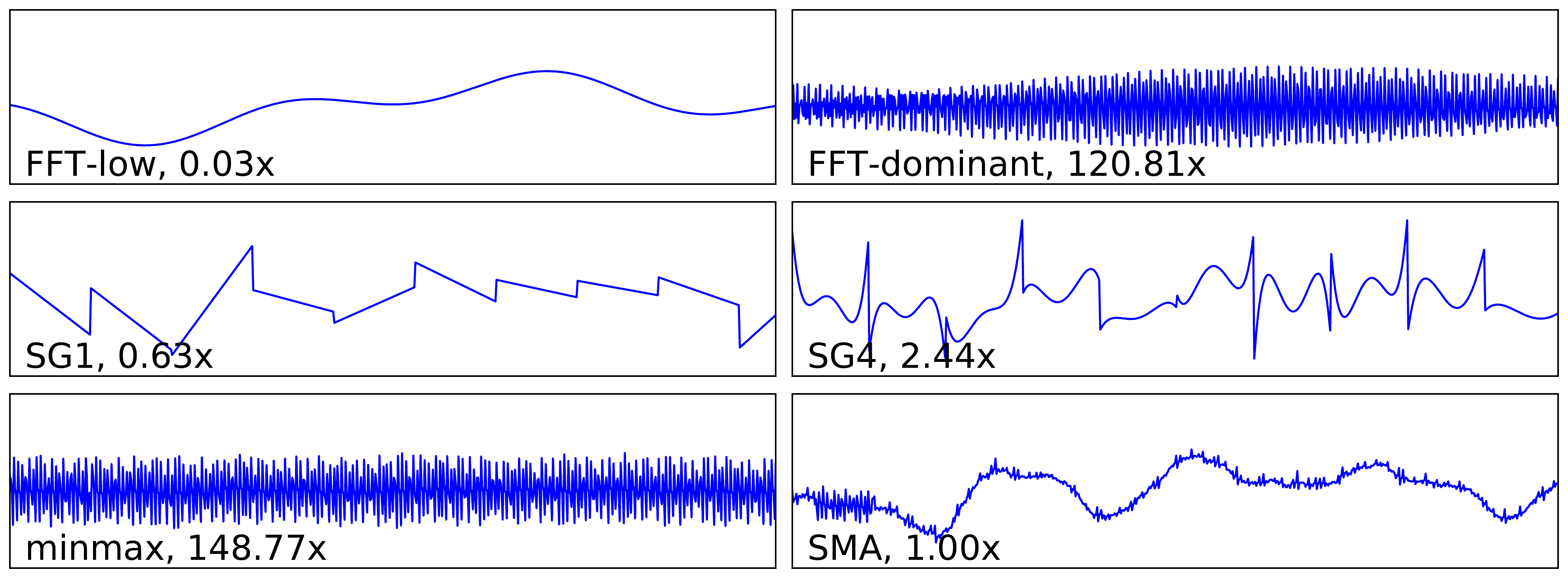}
  \caption{EEG}
  \label{fig:func_36}
\end{subfigure}
\begin{subfigure}[b]{0.9\columnwidth}
  \includegraphics[width=\textwidth]{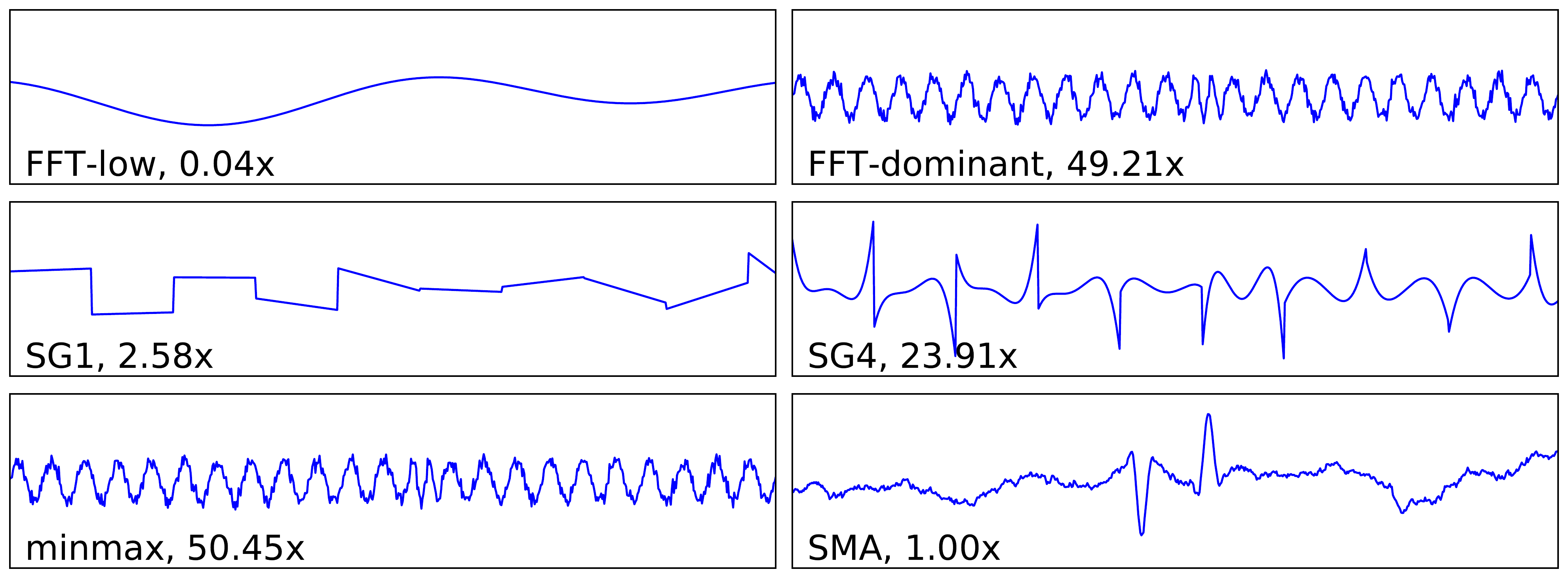}
  \caption{Sine}
  \label{fig:func_38}
\end{subfigure}
\begin{subfigure}[b]{0.9\columnwidth}
  \includegraphics[width=\textwidth]{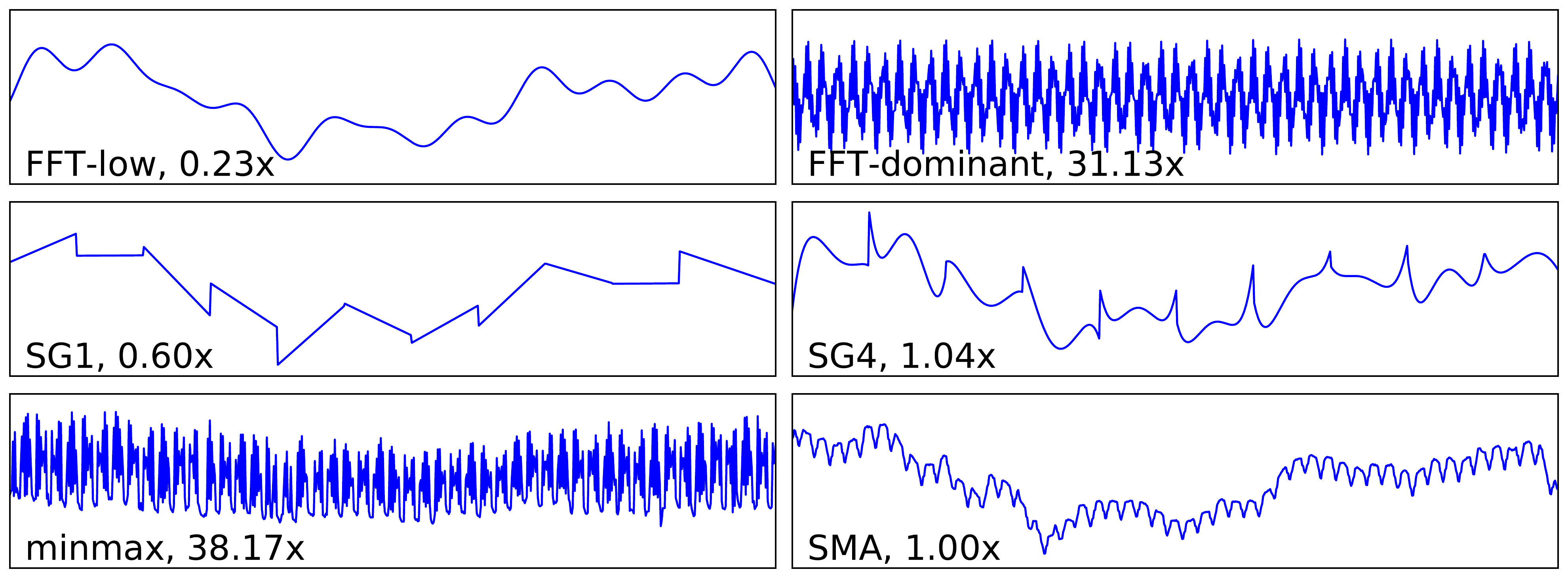}
  \caption{Power}
  \label{fig:func_42}
\end{subfigure}
\caption{Achieved roughness of FFT, Savitzky-Golay filter and Wiener filter over SMA.}
\label{fig:smoothing_func}
\end{figure}

\minihead{Smoothing Function} In addition, we compare achieved roughness of moving average (SMA) with alternative smoothing functions including Fast Fourier transform, Savitzky-Golay filter and the minmax aggregation using the same parameter selection criteria (minimizing roughness subject to kurtosis preservation). We varied the window sizes for Savitzky-Golay and minmax filters, and varied the number frequency components included in the reconstruction for FFTs. Specifically, SG1 approximates data points in a window using a line while SG4 approximates using a polynomial of degree 4; FFT-low reconstructs the signal by composing components in the order of increasing frequency while FFT-dominant composes frequency components of decreasing power. We report achieved roughness compared to SMA for each smoothing function for all user study datasets in Figure~\ref{fig:smoothing_func}. Overall, we found that FFT-dominant and minmax result in high roughness: the former tend to keep the dominant high frequencies in the original time series during reconstruction and the latter, by definition, produces smoothed time series where consecutive points are maximized in distance in the given window. FFT-low, SG1 and SG4, on the other hand, produce smoother plots and occasionally outperform SMA in roughness. Figure~\ref{fig:smoothing_func} contains all smoothed plots for visual comparison.

\section{Additional Visualizations}
\begin{figure}\centering
\includegraphics[width=0.65\linewidth]{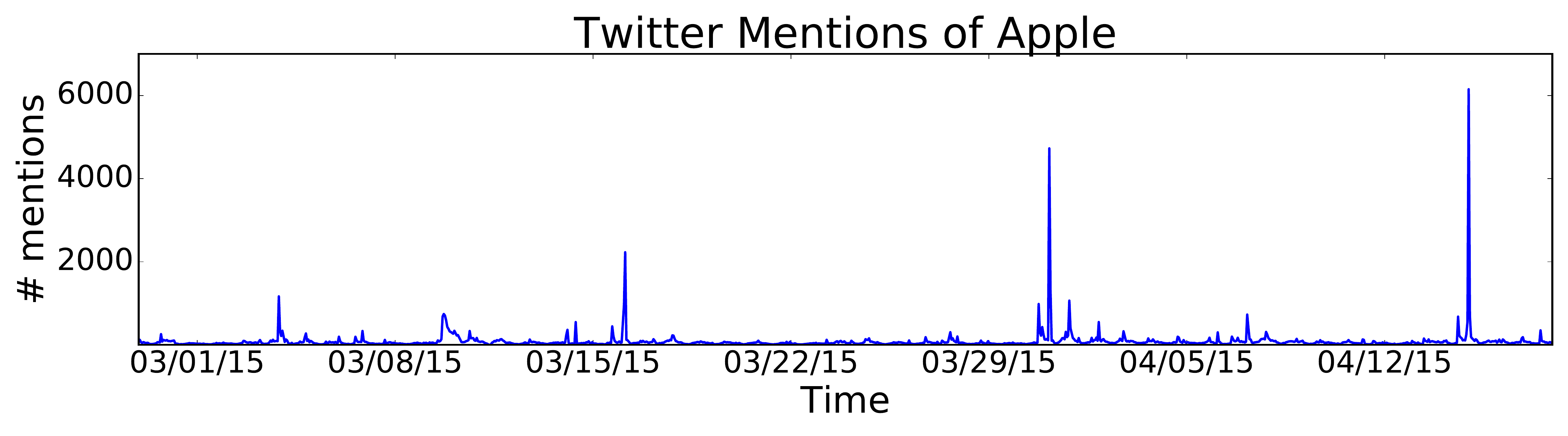}
\caption{Twitter\_AAPL. This plot is left unsmoothed by both exhaustive search and ASAP due to its high initial kurtosis.}
\label{fig:twitter}
\end{figure}

\begin{figure}\centering
\centering
\begin{subfigure}[b]{.65\linewidth}
  \includegraphics[width=\linewidth]{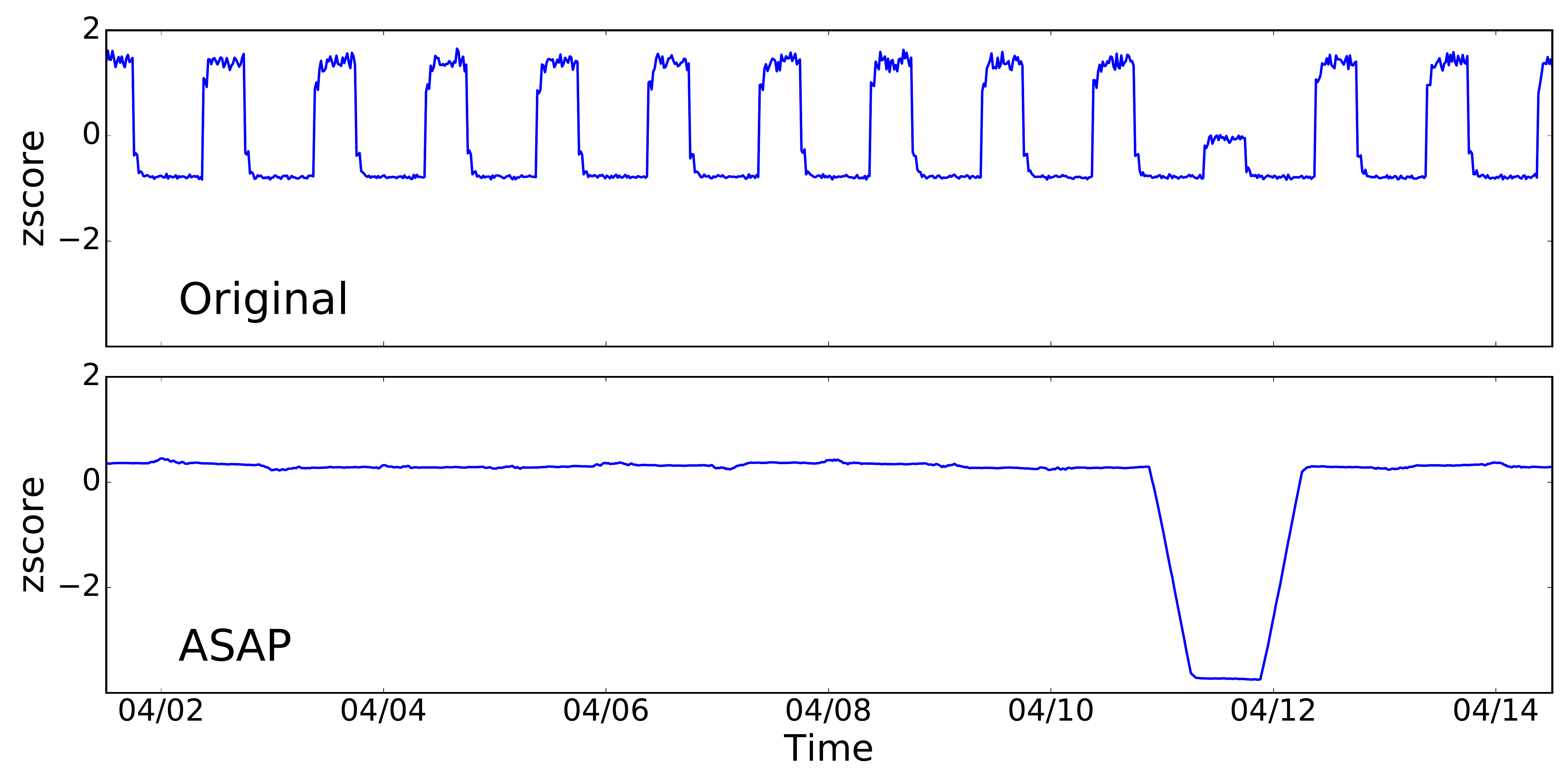}
  \caption{sim\_daily}
  \label{fig:sim-daily}
\end{subfigure}
\begin{subfigure}[b]{.65\linewidth}
  \includegraphics[width=\linewidth]{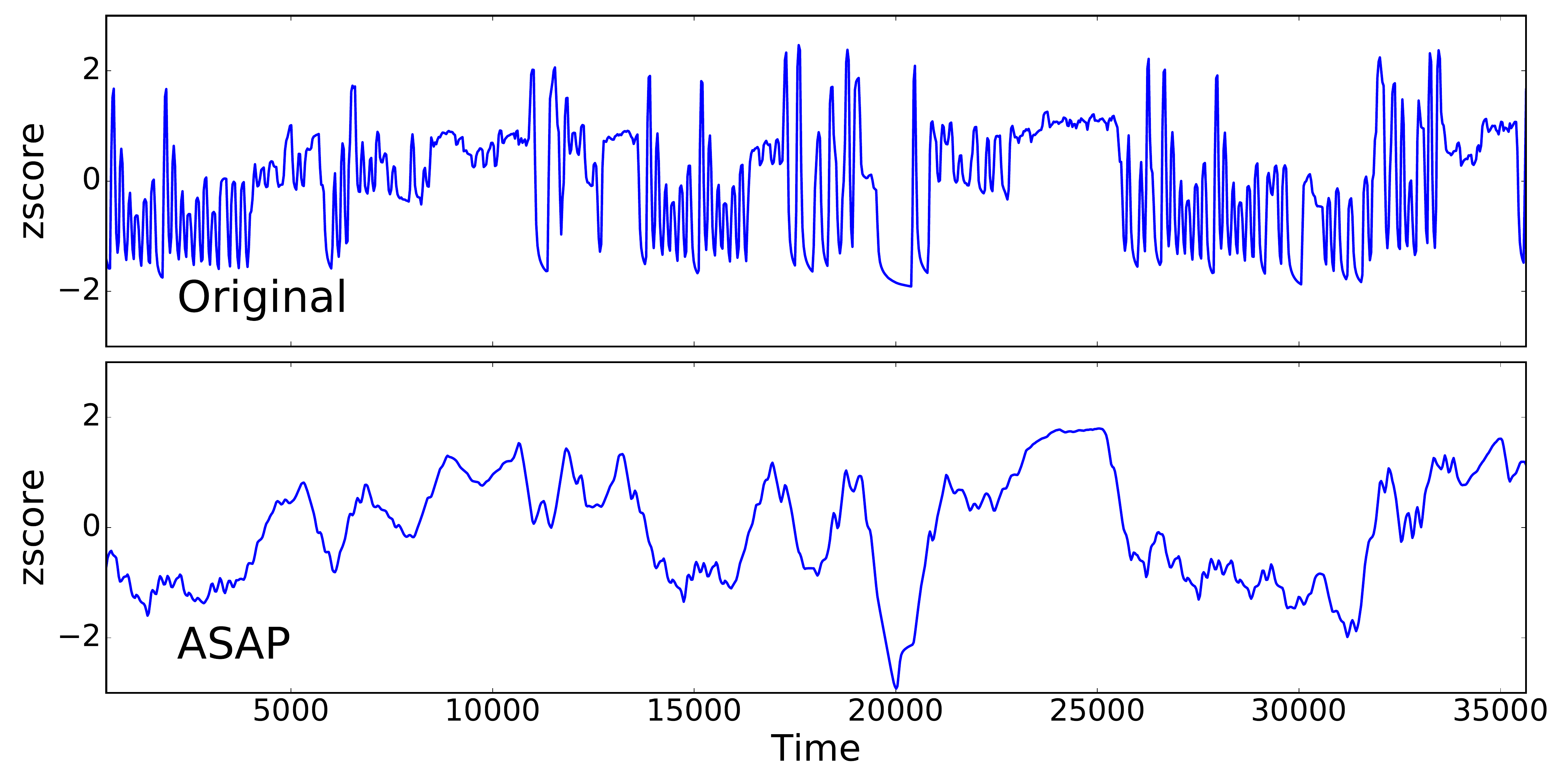}
  \caption{gas\_sensor}
  \label{fig:gas-sensor}
\end{subfigure}
\begin{subfigure}[b]{.65\linewidth}
  \includegraphics[width=\linewidth]{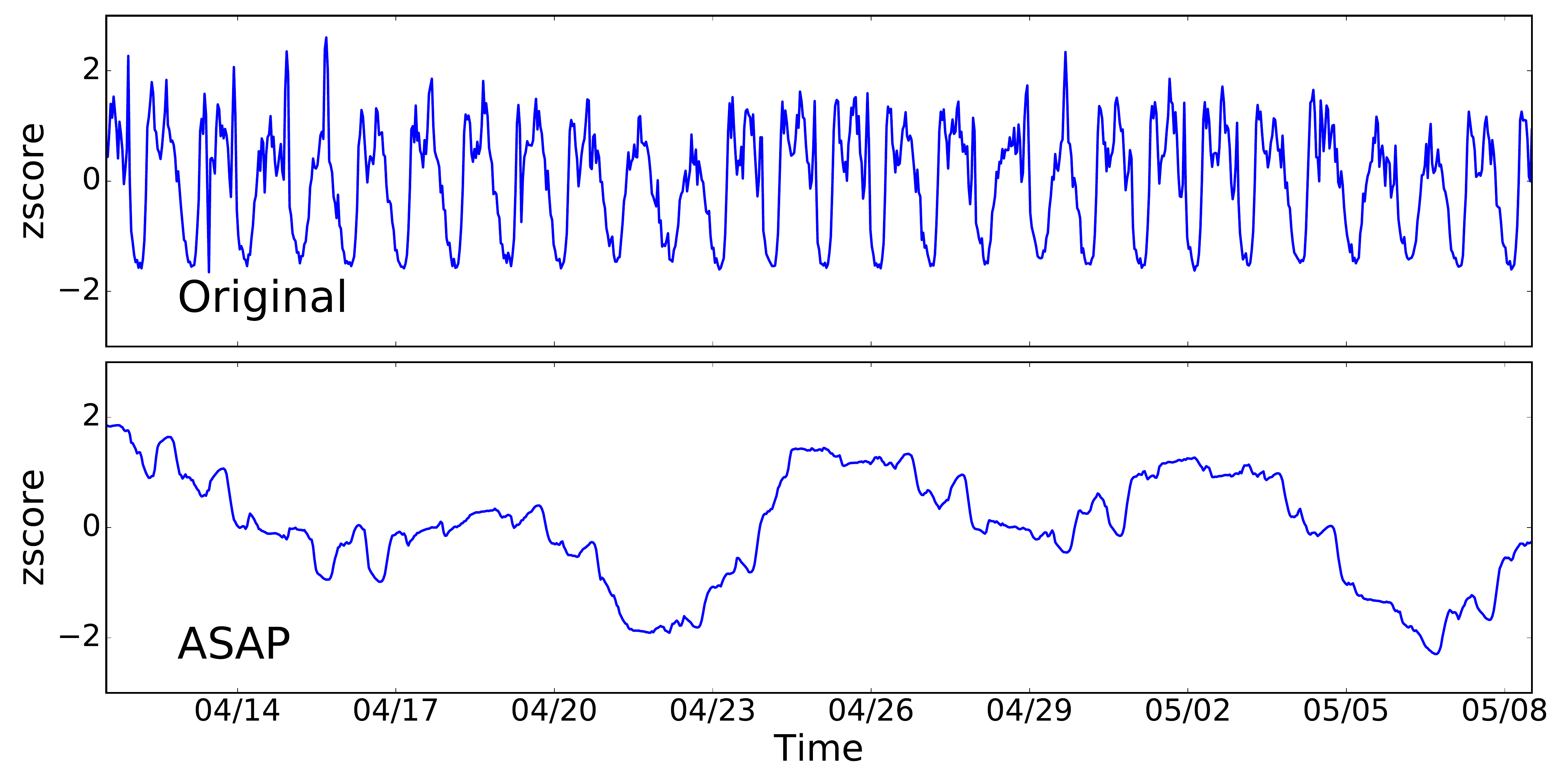}
  \caption{ramp\_traffic}
  \label{fig:ramp-traffic}
\end{subfigure}
\begin{subfigure}[b]{.65\linewidth}
  \includegraphics[width=\linewidth]{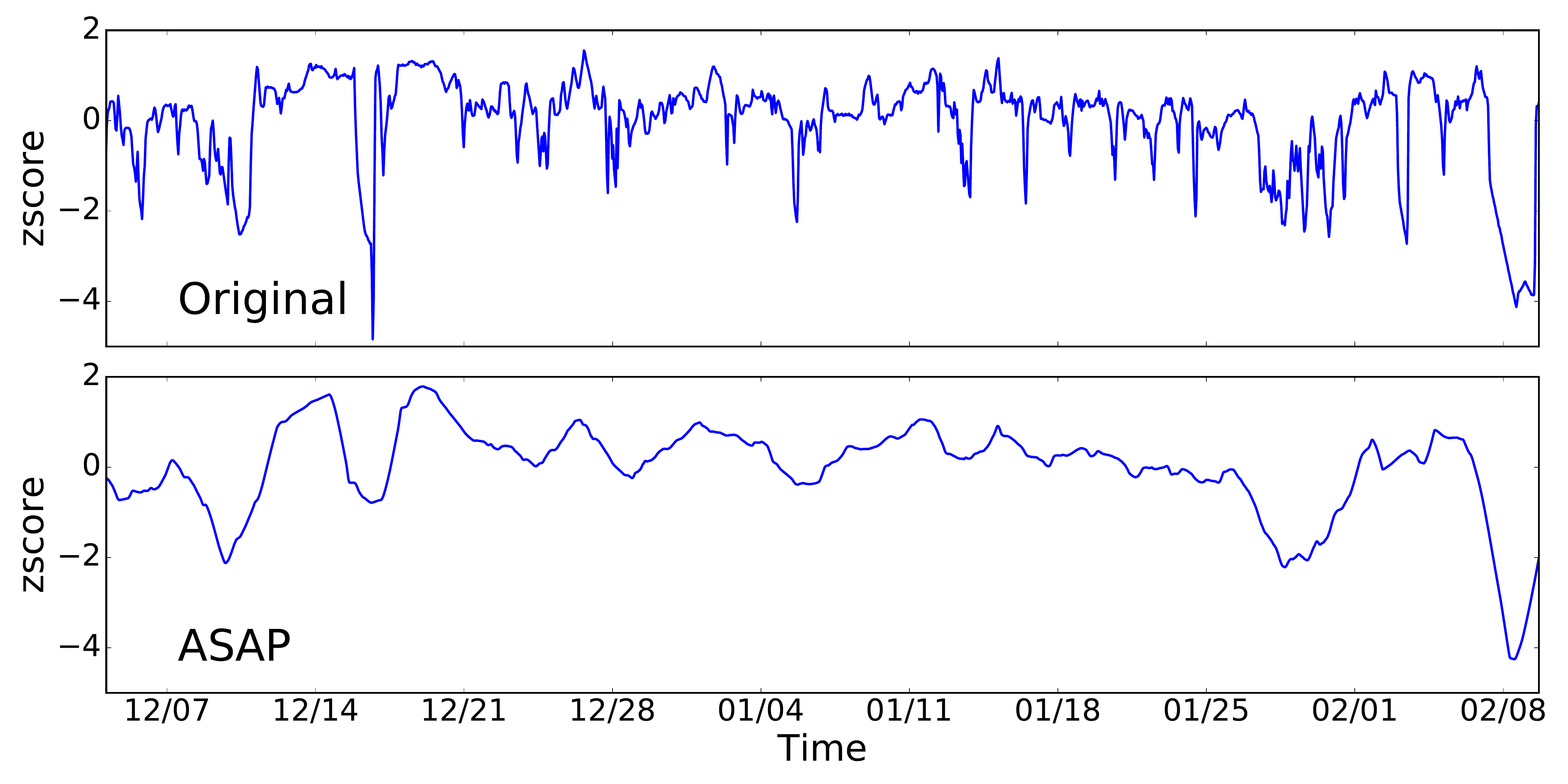}
  \caption{machine\_temp}
  \label{fig:machine-temp}
\end{subfigure}
\begin{subfigure}[b]{.65\linewidth}
	\includegraphics[width=\linewidth]{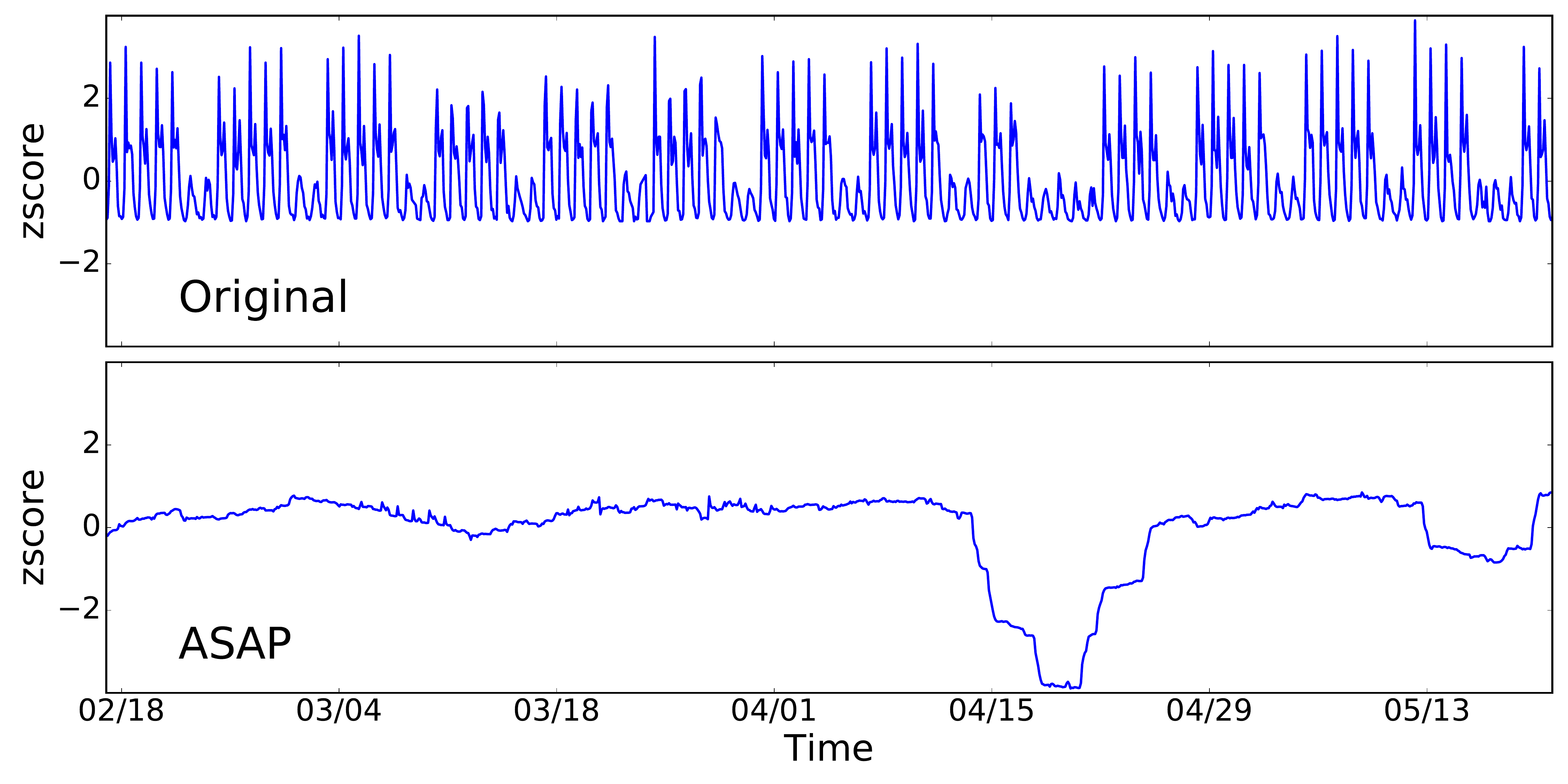}
	\caption{traffic\_data}
\label{fig:traffic-data}
\end{subfigure}
\caption{Original and ASAP-smoothed plots}\label{fig:all-smoothed}
\end{figure}

\begin{figure}\centering
\begin{subfigure}[b]{0.48\columnwidth}
  \includegraphics[width=\textwidth]{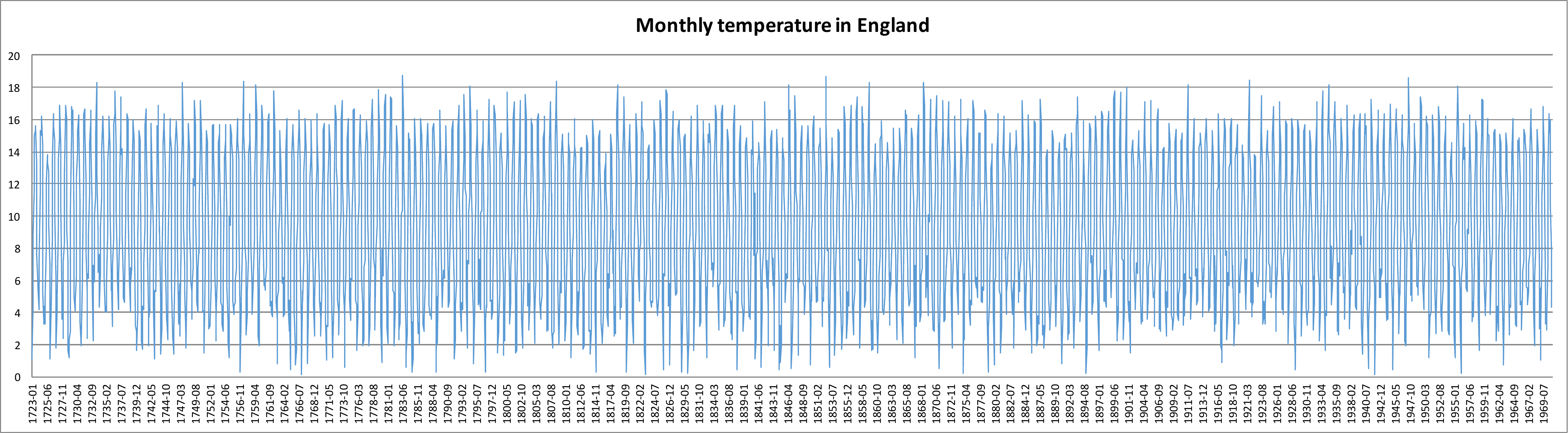}
  \caption{Excel}
  \label{fig:excel}
\end{subfigure}
\begin{subfigure}[b]{0.48\columnwidth}
  \includegraphics[width=\textwidth]{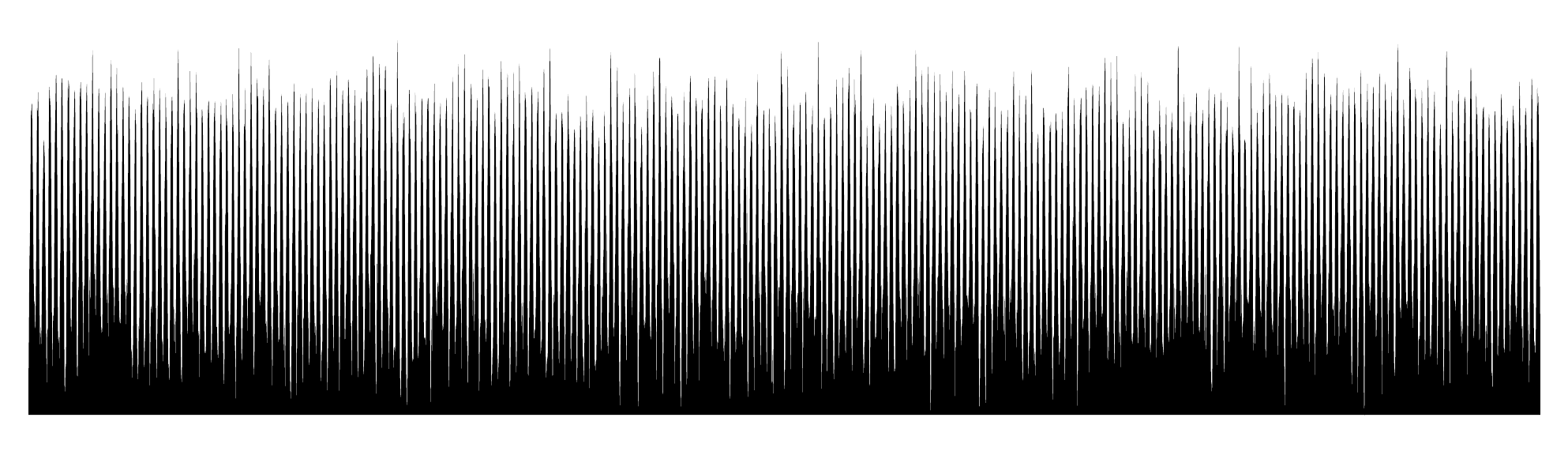}
  \caption{Prometheus}
  \label{fig:prometheus}
\end{subfigure}
\begin{subfigure}[b]{0.48\columnwidth}
  \includegraphics[width=\textwidth,height=0.25\linewidth]{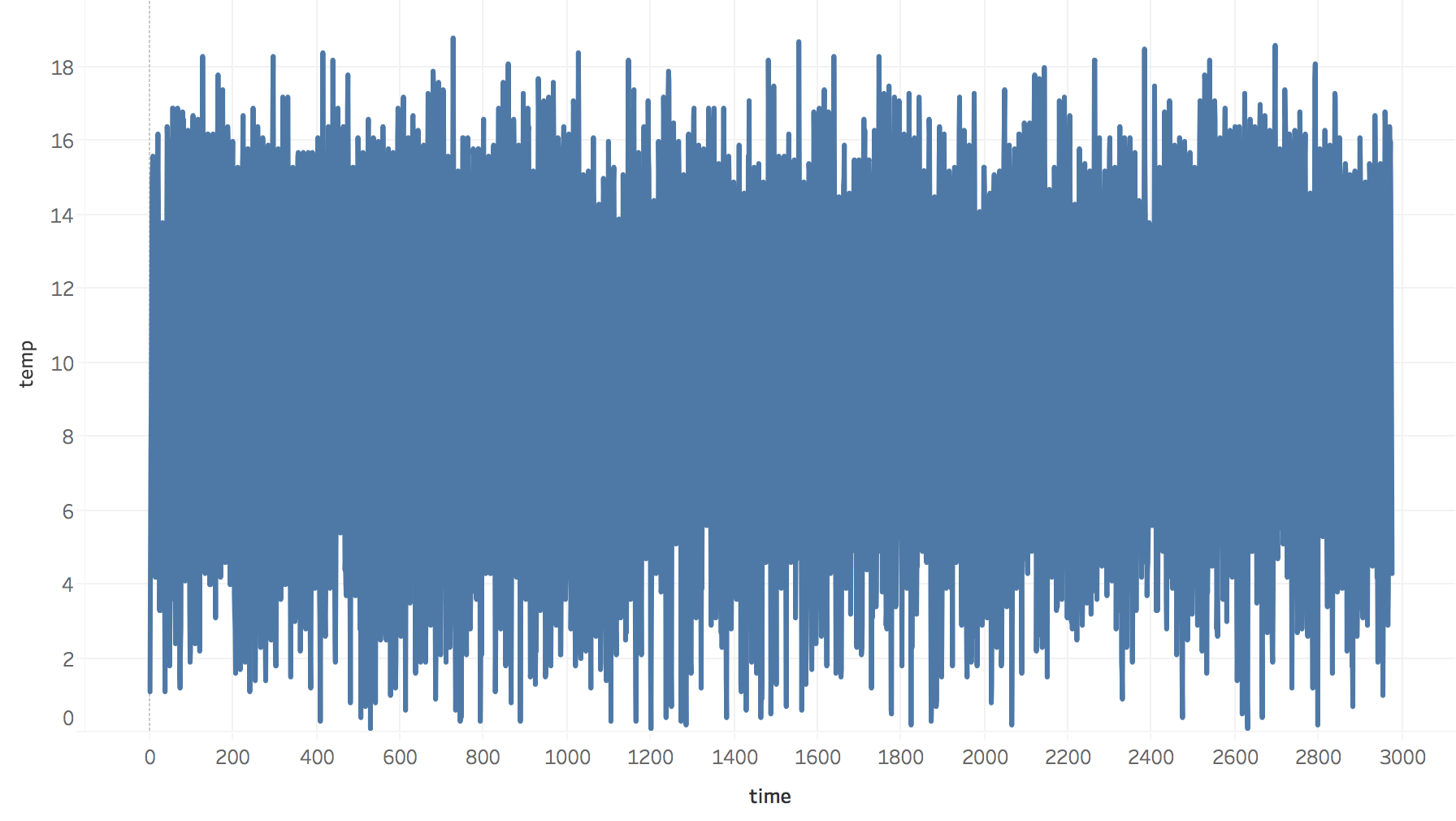}
  \caption{Tableau}
  \label{fig:tableau}
\end{subfigure}
\begin{subfigure}[b]{0.48\columnwidth}
  \includegraphics[width=\textwidth]{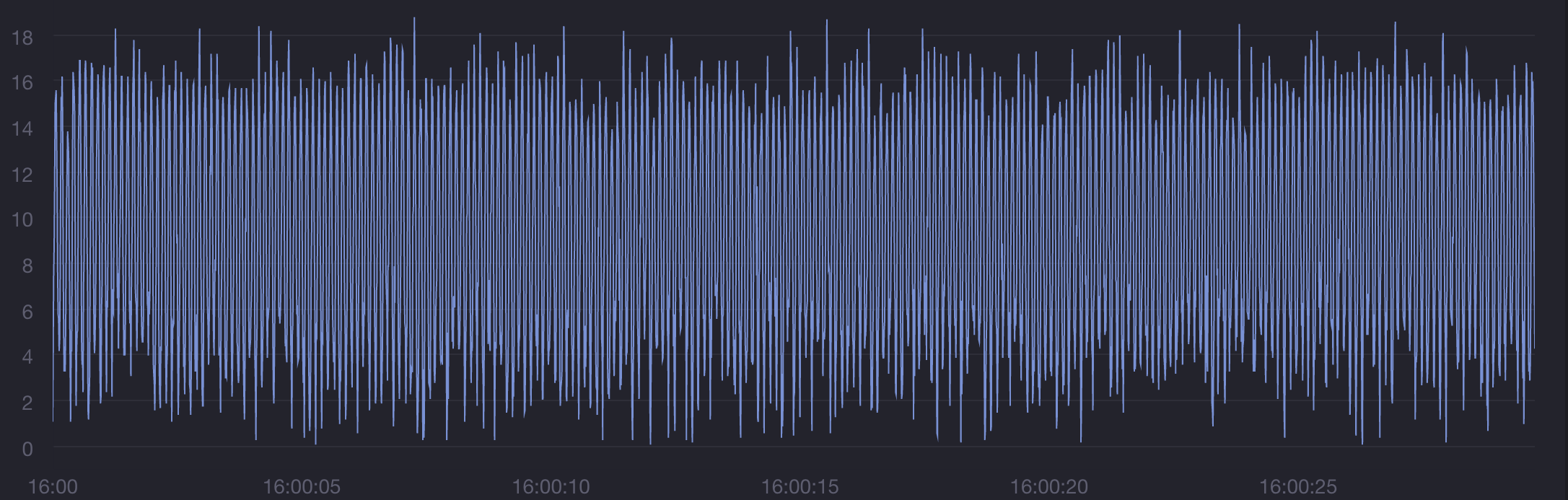}
  \caption{Grafana}
\label{fig:grafana}
\end{subfigure}
\caption{Sample visualizations for the Temp dataset using various existing time series visualization tools; none automatically smoothes out the noise}
\label{fig:tools}
\end{figure}

In this section, we present the remaining plots of raw and (ASAP-)smoothed time series for datasets in Table~\ref{tab:datasets} (Figure~\ref{fig:twitter},~\ref{fig:all-smoothed}), as well as a sample of visualizations generated by commonly used monitoring systems, applications and plotting libraries for the Temp dataset (Figure~\ref{fig:tools}). 

\end{document}